\documentclass[10pt]{article}

\usepackage[T1]{fontenc}
\usepackage{amscd, amsfonts, amsmath, amssymb, amstext, amsthm, caption, epsfig, fancyhdr, float, graphicx, latexsym, mathtools, multicol, multirow, algorithm, chngcntr}
\usepackage{kpfonts}
\usepackage[lofdepth,lotdepth]{subfig}
\usepackage{authblk}
\usepackage[utf8]{inputenc}
\usepackage{bm} 
\usepackage{enumerate}
\usepackage{relsize}
\usepackage{booktabs}
\usepackage[numbers]{natbib}
\usepackage{color}
\usepackage{bm}
\usepackage{parskip}

\usepackage[pdfpagelabels,colorlinks=true,linkcolor=blue,anchorcolor=blue,citecolor=blue,filecolor=blue,menucolor=blue,runcolor=blue,urlcolor=blue]{hyperref}

\usepackage{algpseudocode} 

\usepackage{pgf}
\usepackage[utf8]{inputenc}\DeclareUnicodeCharacter{2212}{-}

\usepackage{caption}
\usepackage[capitalise,nameinlink,noabbrev]{cleveref}
\crefname{equation}{}{}
\crefname{enumi}{}{}

\graphicspath{{Figures/}}

\makeatletter
\newcounter{algorithmicH}
\let\oldalgorithmic\algorithmic
\renewcommand{\algorithmic}{%
  \stepcounter{algorithmicH}
  \oldalgorithmic}
\renewcommand{\theHALG@line}{ALG@line.\thealgorithmicH.\arabic{ALG@line}}
\makeatother

\makeatletter
\DeclareCaptionLabelFormat{andtable}{#1~#2  \&  \tablename~\thetable}
\makeatletter

\makeatletter
\DeclareCaptionLabelFormat{andtable}{#1~#2  \&  \tablename~\bm{\theta}ble}
\makeatletter

\newtheorem{remark}{Remark}[section]
\newtheorem{ex}{Example}

\newtheorem{hp}{Assumption}
\newtheorem{pb}{Problem}

\newcommand*{\UnifDist}{\mathsf{Unif}}

\newcommand*{\GammaDist}{\mathsf{Gamma}}
\newcommand*{\LognormalDist}{\mathsf{LogNorm}}

\newcommand*{\PoissonDist}{\mathsf{Poisson}}

\newcommand*{\cond}{\mid}

\usepackage{xspace}

\newcommand*{\iid}{\textsc{iid}\@\xspace}

\newcommand*{\abc}{{\textsc{abc}}\@\xspace}

\newcommand*{\ess}{\textsc{ess}\@\xspace}

\newcommand*{\kde}{\textsc{kde}\@\xspace}

\newcommand*{\map}{\textsc{map}\@\xspace}
\newcommand*{\bp}{\textsc{bp}\@\xspace}
\newcommand*{\mode}{\textsc{mode}\@\xspace}

\newcommand*{\bt}{\bm{\bm{\theta}}}

\newcommand*{\bx}{\bm{x}}

\newcommand*{\bH}{\bm{H}}

\DeclareMathOperator*{\argmin}{arg\,min}

\newcommand*{\ind}{\mathbb{I}}

\definecolor{MyBlue}{HTML}{1f77b4}
\definecolor{MyGreen}{HTML}{2ca02c}
\definecolor{MyRed}{HTML}{d62728}
\definecolor{MyPurple}{HTML}{9467bd}
\definecolor{MyBrown}{HTML}{8c564b}

\usepackage[percent]{overpic}

\usepackage{float}
\usepackage{microtype}

\renewcommand{\tilde}{\widetilde}

\usepackage{cancel}

\usepackage[marginratio=1:1,height=584pt,width=480pt,tmargin=90pt]{geometry}

\begin{document}

\title{Market-based insurance ratemaking: application to pet insurance}
\author[1]{Pierre-Olivier Goffard \footnote{Email: \href{mailto:goffard@unistra.fr}{goffard@unistra.fr}.}}
\author[2,3]{Pierrick Piette\footnote{Email: \href{mailto:pierrick.piette@gmail.com}{pierrick.piette@gmail.com}.}}
\author[4]{Gareth W. Peters\footnote{Email: \href{mailto:garethpeters@ucsb.edu}{garethpeters@ucsb.edu}.}}
\affil[1]{\footnotesize Université de Strasbourg, Institut de Recherche Mathématique Avancée, Strasbourg, France}
\affil[2]{\footnotesize Univ Lyon, Université Claude Bernard Lyon 1, Institut de Science Financière et d’Assurances (ISFA), Laboratoire SAF EA2429, F-69366, Lyon, France}
\affil[3]{\footnotesize  Hestialytics, Paris}
\affil[4]{\footnotesize  University of California Santa Barbara, Department of Statistics and Applied Probability, Santa Barbara CA 93106-3110, USA}

\maketitle
\vspace{3mm}

\begin{abstract}
This paper introduces a method for pricing insurance policies using market data. The approach is designed for scenarios in which the insurance company seeks to enter a new market, in our case: pet insurance, lacking historical data. The methodology involves an iterative two-step process. First, a suitable parameter is proposed to characterize the underlying risk. Second, the resulting pure premium is linked to the observed commercial premium using an isotonic regression model. To validate the method, comprehensive testing is conducted on synthetic data, followed by its application to a dataset of actual pet insurance rates. To facilitate practical implementation, we have developed an R package called \texttt{IsoPriceR}. By addressing the challenge of pricing insurance policies in the absence of historical data, this method helps enhance pricing strategies in emerging markets.  
\end{abstract}

\emph{MSC 2010}: 62P05, 91G70, 62F15.\\
\emph{Keywords}: Insurance Pricing, Bayesian Inference, Approximate Bayesian Computation, Isotonic Regression.

\section{Introduction} \label{sec:intro}
Modern insurance pricing relies on predictive modeling methods to ensure that premiums reflect, as accurately as possible, the average cost of claims. To achieve this, insurers rely on historical data to train statistical learning models and calculate what is called the pure premium. Although the foundation of standard actuarial practice often rests on generalized linear models (GLM), see \citet{renshaw_1994}, the relentless evolution of data science has ushered in a new era where more sophisticated machine learning algorithms are also coming into play, see \citet{BlierWong2020} and the reference therein. However, a challenge arises when an insurance company enters a new market, lacking historical data on the risks it aims to cover. In this context, conventional predictive modeling tools are failing, leaving insurers at a crossroad looking for innovative solutions to navigate uncharted territory.

\noindent Although an insurance company may lack historical data in a new market, there is an attractive alternative: Collect and analyze market data consisting of rates offered by competitors for similar insurance policies. Our approach leverages these market data to provide insights into the underwritten risk leading to the calculation of insurance premiums. Our objective is to develop a methodology that determines suitable commercial premiums based on the observed commercial rates of competitors.

\noindent The data collection process involves obtaining insurance quotes.  To gather these quotes, one can either visit insurance company websites and answer several questions about the insured risk or obtain survey data responses from a data broker that has automated such a process. The premiums quoted depend on the responses provided by the customer. In this paper, we focus on a pet insurance application, but we have opted for generic notation because we believe that the method could be suitable for other insurance products. Pet insurance covers veterinary expenses for pets, including unexpected injuries (e.g., foreign object ingestion, broken bones), illnesses (e.g., cancer, glaucoma, hip dysplasia, parvovirus), surgeries (e.g., cruciate ligament tears, cataracts), medications, diagnostic tests (e.g., X-rays, blood tests, MRIs), and emergency exam fees. However, most pet insurance policies exclude coverage for pre-existing conditions, routine and preventive care, spaying/neutering, vaccinations, and other specific exclusions. To determine the cost of coverage, pet owners must provide details about their pet’s species, breed, age, and gender.  These characteristics are referred to as rating factors in actuarial science and are crucial for risk classification. For an overview of this topic, we direct the reader to the work of \citet{Antonio2011}. An insurance company looking to enter a new market would naturally identify these risk factors when collecting data on the premiums offered by competitors.

Within a specific risk class, the risk is represented by a positive random variable, denoted as \(X\), which quantifies the total amount of the claims during the insurance policy period. Insurance companies mitigate this risk by offering coverage for a portion of \(X\), denoted as \(g(X) < X\), in exchange for a premium. The process involves calculating the pure premium, defined as
\(p = \mathbb{E}[g(X)].\) Customers are then presented with a commercial premium derived from the pure premium as \(\Tilde{p} = f(p) > p,\) where \(f\) represents the loading function. Our problem is formulated as follows: given a collection of insurance quotes \(\mathcal{D} = \{\Tilde{p}_1,\ldots, \Tilde{p}_n\}\) corresponding to a specific risk, with variations in loading and coverage functions, represented as
\[
\Tilde{p}_i = f_i\{\mathbb{E}[g_i(X)]\},\text{ } i = 1,\ldots, n,
\]
we aim to study the distribution of the underlying risk $X$ and approximate the loading functions in  order to price our own insurance policies relative to current market premia and risk loadings.\\

\noindent Our approach assumes that the distribution of the risk \(X\) is parameterized by \(\theta \in \Theta\subset\mathbb{R}^d\). Ideally, if the loading functions \(f_i\) were known for $i = 1,\ldots, n$, a procedure similar to the generalized method of moments could be applied (see \citet{Hansen1982}). Unfortunately, these functions are unknown in this context to anyone not internal to the companies from which the premium quotes were obtained. Therefore, in order to proceed we will perform an estimation under a two-stage procedure. We posit a prior distribution $\pi(\theta)$ to delimit the parameter space of the risk model from which we can readily sample \(\theta\).  We compute the pure premiums \(p_i^\theta = \mathbb{E}_\theta[g_i(X)]\),  based on the known coverages \(g_i\) provided by the insurance policy.  If $X$ follows a compound distribution (i.e. a random sum) then the pure premium does not have a closed form expression but can be approximated to any desired level of precision using simulations.
 
In the second stage,  the loading functions \(f_i: \mathbb{R}_+ \to \mathbb{R}_+\) are approximated using an isotonic regression model, chosen for its ability to maintain the monotonic relationship between pure and commercial premiums—a desirable feature. Additionally, market data is inherently noisy, and isotonic regression provides robustness to outliers, superior to that of simple linear regression, see further discussion on estimation properties of isotonic regression in \cite{luss2014generalized} and references therein.  The procedure may be summarized as follows:

\begin{enumerate}
  \item Sample a parameter value $\theta$,
  \item Compute the pure premiums $p_i^{\theta}$ for each of the insurance policies $i = 1,\ldots, n$,
  \item Fit an isotonic function $f$ to learn the relationship between the commercial premium $\Tilde{p}_i$ and the pure premium $p_i^{\theta}$,
  \item Build the 'synthetic' market data $\mathcal{D}^{\theta}$ by applying the estimated loading $f$ to the pure premium $f(p_i^\theta)$ for $i=1,\ldots, n$,
  \item If the observed and synthetic market data are close enough,  according to a prespecified distance,  then we store the parameter value $\theta$ and the associated loading function $f$.
\end{enumerate}
After iterating the above steps, we get a sequence of parameter-loading function pairs: $(\theta_1,f_1),(\theta_2,f_2 ), \ldots$. This sequence allows us to price our own insurance policies. The problem we tackle is an inverse problem and our solution is inspired from indirect inference methodologies pionneered by \citet{Gourieroux1993}. The proposed algorithm to search the parameter space resembles Approximate Bayesian Computation (ABC) algorithms described in the book of \citet{SiFaBe18}. The parameter values sampled $\theta_1, \theta_2,\ldots$ by the algorithm yields an approximate posterior distribution $\pi(\theta|\mathcal{D})$. This posterior distribution accounts for the uncertainty around the estimated parameter value due to the use of Monte Carlo simulation to calculate the pure premium. ABC algorithms have found successful applications in a range of actuarial science and risk management problems. We refer the readers to the works of \citet{peters2010chain,dean2014parameter,peters2006bayesian} and \citet{Goffard2021} for further insights. Isotonic regression, a well-established statistical methodology, see for instance \citet{barlow1972statistical}, plays a central role in our approach. A recent application in actuarial science addresses the autocalibration challenges that can arise when pricing insurance contracts using machine learning algorithms, see the work of \citet{Wuethrich2023}.

The rest of the paper is organized as follows. \cref{sec:model_set_up} describes the risk model used in this study and discusses insurance pricing principles. \cref{sec:methodology} provides a detailed account of the algorithmic procedure. Our method is presented as an Approximate Bayesian Computation (ABC)-type optimization algorithm, which incorporates a simple isotonic regression model. \cref{sec:simulation} presents the results of a simulation study designed to showcase the performance of our method in a controlled environment. Then, we apply our algorithm on a dataset made of real-world pet insurance rates in \cref{sec:petApplication}.  Finally, we conclude in \cref{sec:conclusion} and discuss the perspectives and limits of our methodology for potential applications on other insurance products.

\section{Model set up and insurance premium computation}\label{sec:model_set_up}
An individual seeks to hedge against a risk $X$ modeled by a positive random variable, over a given period of time, say one year. A common model used for $X$ in property and casualty insurance is given by a compound loss variable 
\begin{equation}\label{eq:collective_model}
X = \sum_{k=1}^{N}U_k,
\end{equation}
where $N$ is a counting random variable and the \(U_k\)'s are independent and identically distributed (\iid) positive random variables independent from $N$. The random variable $N$ is the number of occurrences of an event over a given time period (annually), each of these events is associated to a compensation $U_k$.  We assume here that $X$ represents a risk that belongs a specific risk class determined by risk factors.

\subsection{Pure premium computation}\label{ssec:pp}

\noindent An insurance company offers to bear part of this risk $g(X)\leq X$ in exchange for a premium which should compensate the average cost of claim given by \(p = \mathbb{E}\left[g(X)\right],\) referred to as the pure premium. We consider in this work a function $g$ defined as 
$$
g(x) = \min(\max(r\cdot x-d,\,0),\, l), 
$$ 
where $r\in(0,1]$ is the coverage rate, $d>0$ is the deductible and $l>0$ is the limit. We illustrate the impact of the parameters of the insurance coverage in \cref{ex:insurance_coverage}.
\begin{ex}\label{ex:insurance_coverage}
Let us consider a scenario where the risk has a Poisson-lognormal distribution $X\sim\PoissonDist(\lambda = 3)-\LognormalDist(\mu = 0, \sigma = 1)$ and that $n = 100$ insurance coverages are proposed. These are characterized by a rate, a deductible and a limit, set randomly as 
$$
r_i\sim\UnifDist([0.5, 1])\text{, }d_i\sim\UnifDist([0.5, 6])\text{, and }l_i = \infty\text{, for }i = 1, \ldots, 100.
$$
\cref{fig:pp_plot} shows the pure premiums
\begin{equation}\label{eq:pps}
p_i = \mathbb{E}(g_i(X)) = \mathbb{E}(\min(\max(r_i\cdot x-d_i,\,0),\, l_i)),\text{ }i = 1,\ldots, 100.
\end{equation}
as a function of the rates and deductibles.
\begin{figure}[!ht]
\centering
  \includegraphics[width=0.5\linewidth]{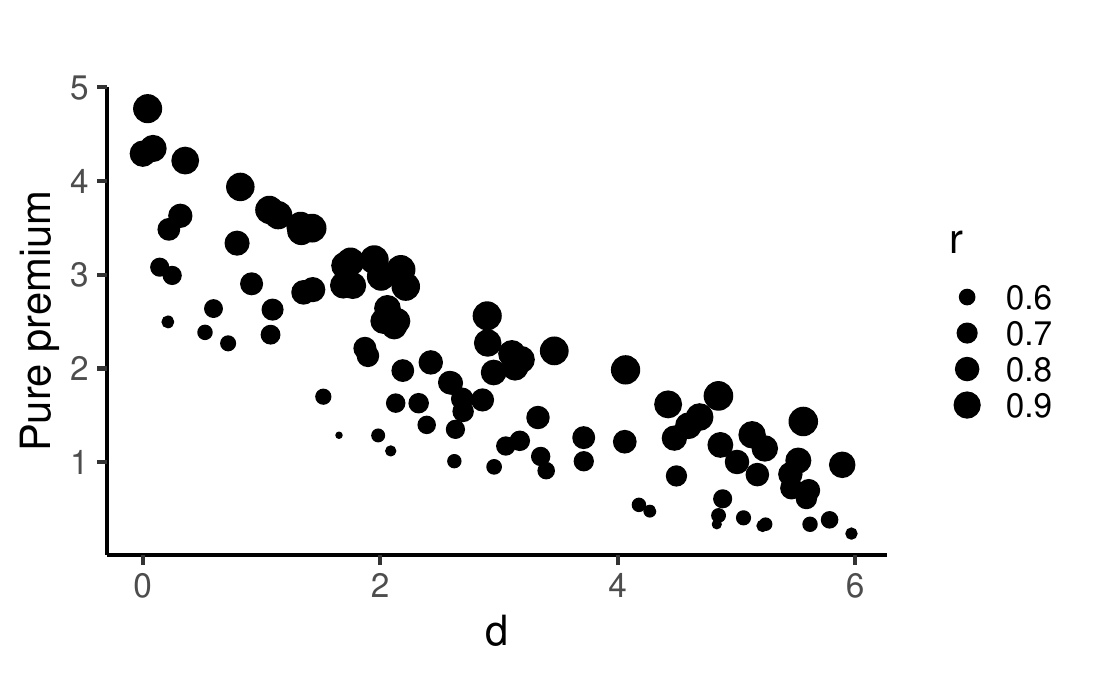}
  \caption{Pure premiums as a function of the rate of coverage (r) and the deductible (d) for a $\PoissonDist(\lambda = 3)-\LognormalDist(\mu = 0, \sigma = 1)$ risk.}
  \label{fig:pp_plot}
\end{figure}

The pure premium are increasing in the rates and decreasing in the deductible. Note that the pure premiums were estimated via simulations to overcome the lack of explicit formula for the distribution functions of $X$. 
\end{ex}

In practice, the rate offered to policyholders includes a loading to compensate for the variability of the risks and to cover the management costs. We describe this loading in the next section.
\subsection{From pure premiums to commercial premiums}\label{ssec:cp}
Let $f:\mathbb{R}_+\mapsto\mathbb{R}_+$ be a non-decreasing function, such that
$$
\Tilde{p} = f(p)\geq p.
$$
The function $f$ is referred to as the loading function. As the commercial premium is a function of the pure premium then we are applying the expectation premium principle. Other premium principles are also possible like the standard deviation principle discussed in \cref{app:other_premium_principle}. A simple loading function is linear in the pure premium as 
$$
f(x) = (1+\eta)x,
$$
where $\eta>0$. The loading functions used by insurance companies are unknown to us and vary from one insurance company to the other. We follow up on \cref{ex:insurance_coverage} in \cref{ex:safety_loadings} where we randomize the linear link between pure and commercial premium.
\begin{ex}\label{ex:safety_loadings}
Take the pure premiums of \cref{ex:insurance_coverage} and apply the following linear loadings
$$
\eta_i \sim\UnifDist([0.5, 2]),\text{ for }i=1,\ldots, n. 
$$ 
The commercial premium then relates to the pure premium as
\begin{equation}\label{eq:cps}
\Tilde{p}_i = (1+\eta_i)p_i,\text{ for }i=1,\ldots, n. 
\end{equation}
\cref{fig:cp_plot} displays the commercial premium as a function of the pure premium. 
\begin{figure}[!ht]
\centering
  \includegraphics[width=0.7\linewidth]{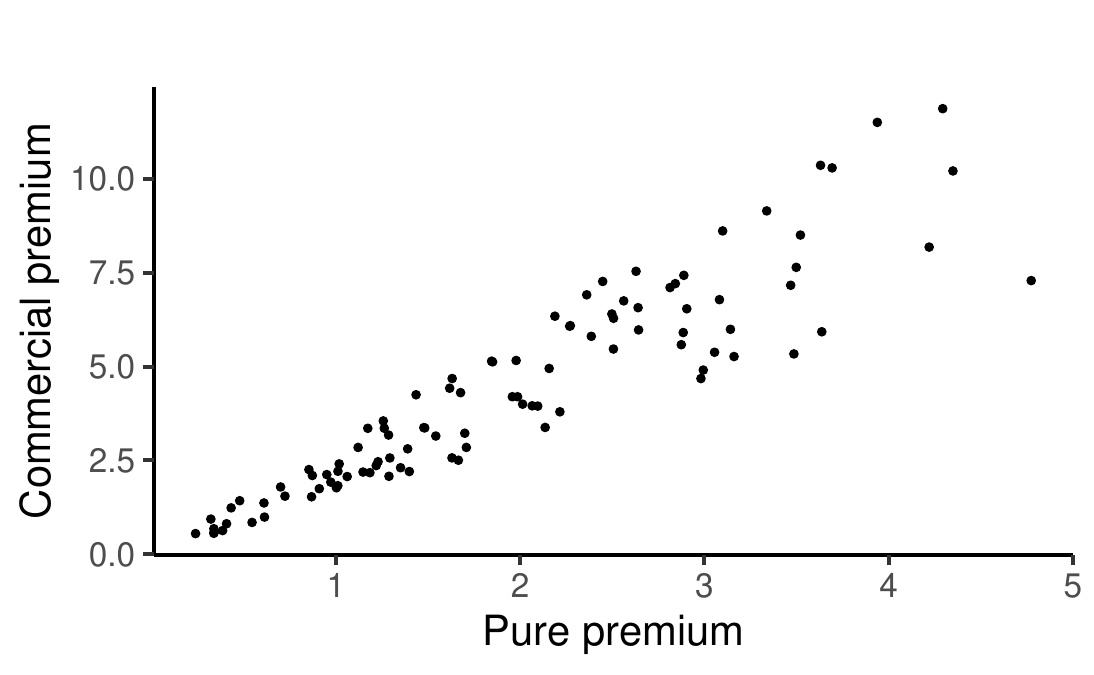}
  \caption{Pure premium as a function of the commercial premium offered by various insurance companies.}
  \label{fig:cp_plot}
\end{figure}

\end{ex}

We only observe the commercial premium $\Tilde{p}_1,\ldots, \Tilde{p}_n$ and we would like to learn from them about the risk $X$ and the loading functions $f_1,\ldots, f_n$. We formulate our problem and describe our solution in the next section.

\section{Market derived insurance ratemaking}\label{sec:methodology}

The data at hand is a collection of insurance rates $\Tilde{p}_{1:n} = \{\Tilde{p}_1,\ldots, \Tilde{p}_n\}$ associated to insurance coverages $g_1,\ldots, g_n$ for a given risk $X$ within a particular risk class. We suppose that these insurance rates $\mathcal{D}$ were derived according to the model described in \cref{hp:model_assumption}.
\begin{hp}\label{hp:model_assumption}
The commercial rate $\Tilde{p}_{1:n} = \{\Tilde{p}_1,\ldots, \Tilde{p}_n\}$ are given by

$$
\Tilde{p}_i = f_i(p_i) = f_i\left\{\mathbb{E}\left[g_{i}(X)\right]\right\} 
,\text{ }i = 1,\ldots, n,
$$

where the risk $X$ is a random variable defined in \eqref{eq:collective_model}. The loading functions $f_1,\ldots, f_n$ are unknown. The insurance coverage functions $g_{1},\ldots, g_n$ are known and of the form
\[
g_i(x) = \min(\max(r_i\cdot x-d_i,\,0),\, l_i)\text{, for }i = 1,\ldots, n.
\] 
\end{hp}

Suppose that the distribution of $X$ is parametrized by an unknown parameter $\theta\in\Theta\subset\mathbb{R}^d$. We wish to  find the parameters $\theta$  that best explain our data based on a loading function $f$ that attempts to average out the loading functions $f_1,\ldots, f_n$ used by the competitors. Our approach alternates between proposing values for the parameters $\theta$ to calculate the pure premiums and subsequently linking these to the commercial premiums. \cref{ssec:problem_1} discusses a "simpler" optimization problem in which we do not have to handle the loading functions. We use this preliminary problem to introduce our distance function and the notion of model identifiability. \cref{ssec:optimization_problem_real} considers the actual problem involving the commercial premiums. The link between pure and commercial premiums is approximated using an isotonic regression model. Regularization terms are included in the distance used to compare observed and model-generated rates to mitigate the identifiability issue. Finally, \cref{ssec:algo} refines the accept-reject algorithm laid out in the introduction to search the parameter space in a more efficient way.

\subsection{Optimization problem based on the pure premium}\label{ssec:problem_1}
Assume that  we hold a collection of pure premiums  $p_{1:n}$ and consider the following optimization problem:
\begin{pb}\label{pb:optimization_problem_simple}
Find $\theta\in \Theta\subset\mathbb{R}^d$ to minimize \(d\left(p_{1:n}, p_{1:n}^\theta\right)\), where 
$$
p_i^\theta =\mathbb{E}_\theta\left[g_{i}(X)\right],\text{ for }i = 1,\ldots, n, 
$$
are the pure premium associated to the risk $X$ parametrized by $\theta$ and $d(\cdot, \cdot)$ denotes a distance function over the observation space.
\end{pb}
We measure the discrepancy between observed and model-generated pure premiums using the root mean square error (RMSE) defined as
\begin{equation}\label{eq:sse_distance_pure}
\text{RMSE}\left(p_{1:n}, p^\theta_{1:n}\right) = \sqrt{\sum_{i=1}^nw^{\text{RMSE}}_i \left(p_i - p_i^\theta\right)^2},
\end{equation}
for a candidate risk parameter $\theta$. The weights $w^{\text{RMSE}}_i > 0$ for $i = 1, \ldots, n$ allow us to place greater emphasis on specific data points. The statistical framework is that of minimum distance estimation. We do not have access to the full shape of the data distribution. We must base our inference on specific moments, just as in the generalized method of moments, a popular method among econometricians (see \citet{Hansen1982}).The model is identifiable if there exists a unique estimator $\theta^\ast$ such that 
\begin{equation}\label{eq:optimization_knowing_pure_premium}
\theta^\ast  = \underset{\theta\in \Theta}{\argmin}\,\text{RMSE}\left(p_{1:n}, p^\theta_{1:n}\right).
\end{equation}
Existence stems from the fact that the parameter space $\Theta$ is compact and the map $\theta\mapsto\text{RMSE}(p_{1:n}, p_{1:n}^\theta)$ is continuous. Uniqueness is more difficult to verify as it depends on the functional $g_i$'s. Given the model for $X$ and the insurance coverages, the pure premium does not have an analytical expression making it difficult to show the convexity of \eqref{eq:sse_distance_pure}. A simple necessary condition is that the number of parameters must be smaller than $n$, the number of moments considered. The shape of our insurance coverage function suggests that we may successfully identify the correct parameters, as demonstrated in \cref{ex:model_identifiability_pp}.
\begin{ex}\label{ex:model_identifiability_pp}
We consider the same model as in \cref{ex:insurance_coverage}. Recall that the risk has a Poisson-lognormal distribution $X\sim\PoissonDist(\lambda = 3)-\LognormalDist(\mu = 0, \sigma = 1)$ and that $n = 100$ insurance coverages are proposed. The rates, deductibles and limit,  are set randomly as 
$$
r_i\sim\UnifDist([0.5, 1])\text{, }d_i\sim\UnifDist([0.5, 6])\text{, and }l = \infty\text{, for }i = 1, \ldots, 100.
$$
Let us assume that $\mu =0$ and compute the pure premium over a grid of values for $\lambda$ and $\sigma$. \cref{fig:pp_contour_plot} shows the $\text{RMSE}[p_{1:n}, p_{1:n}^\theta]$ depending on the value of $\lambda$ and $\sigma$ for $(\lambda,\sigma)\in [0,5]\times  [0,2]$.
\begin{figure}[!ht]
\centering
  \includegraphics[width=0.5\linewidth]{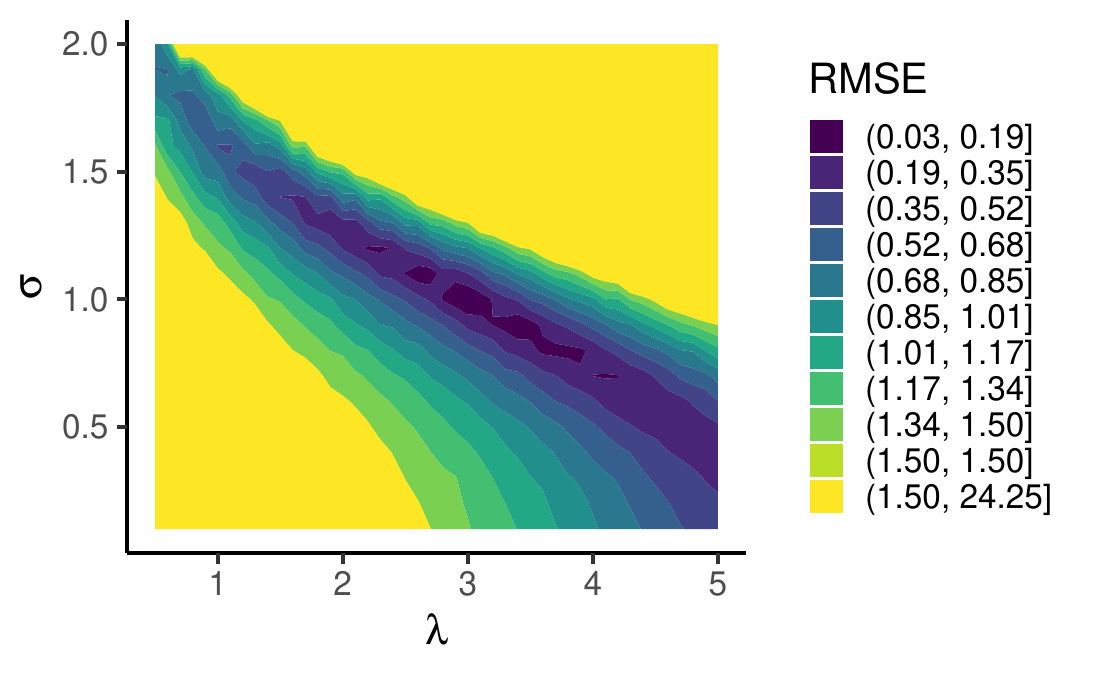}
  \caption{Contour plot of $\text{RMSE}(p_{1:n}, p_{1:n}^\theta)$ for $\mu = 0$ and $(\lambda,\sigma)\in [0,5]\times [0,2]$.}
  \label{fig:pp_contour_plot}
\end{figure}
This contour plot shows minimal RMSE values around the true parameter values \((\lambda_0, \sigma_0) = (3, 1)\).
\end{ex}

If uniqueness of the solution cannot be verified, a workarround consists in adding regularization terms to the discrepancy measure. We explore this dircetion in \cref{ssec:optimization_problem_real} where the actual problem is treated. The issue is further addressed by uing a particle based optimization methods to search the parameter space and provide a set of admissible candidate parameters. Such an optimization method is presented in \cref{ssec:algo}. This identifiability issue from an empirical point of view in the online supplementary material\footnote{\url{https://github.com/LaGauffre/market_based_insurance_ratemaking/blob/main/latex/supp_material.pdf}}.

\subsection{Optimization problem based on the commercial premiums}\label{ssec:optimization_problem_real}
Assume that we hold a set of  commercial rates $\Tilde{p}_{1:n} = \{\Tilde{p}_{1},\ldots, \Tilde{p}_{n}\}$ defined as

$$
\Tilde{p}_i = f_i(p_i) = f_i\left\{\mathbb{E}\left[g_{i}(X)\right]\right\},\text{ }i = 1,\ldots, n.
$$

and consider the following optrimization problem. 
\begin{pb}\label{pb:optimization_problem}
Find $\theta\in \Theta\subset\mathbb{R}^d$ and $f:\mathbb{R}_+\mapsto \mathbb{R}_+$ to minimize \(d\left[\Tilde{p}_{1:n}, f\left(p_{1:n}^\theta\right)\right]\), where the function $f$ is applied elementwise on $p_{1:n}^\theta$ and $d(\cdot, \cdot)$ denotes a distance function over the observation space, subject to 
\begin{equation}\label{eq:constraint}
\Tilde{p}_i \geq p_i^\theta,\text{ and } f(p_i^\theta) \geq p_i^\theta, \text{for }i = 1,\ldots, n.
\end{equation} 
\end{pb}
Our first task is to find a generic function $f$ to represent the safety loading functions $f_i$'s used by the competitors. For this, we use isotonic regression. It is a statistical technique used for fitting a non-decreasing function to a set of data points. The idea is that if two  pure premiums satisfy $p_i\leq p_j$ then the commercial premium should also verify $\Tilde{p}_i\leq \Tilde{p}_j$. Consider a collection of candidate pure premiums $p_{1:n}^\theta$, associated to a candidate estimate of the risk parameter $\theta$.  Our data points are therefore pairs of pure and commercial premiums $(p_i^\theta,\Tilde{p}_i)_{i=1,\ldots, n}$. Suppose the pure premiums have been ordered such that $p_i^\theta\leq p_j^\theta$ for $i\leq j$, isotonic regression seeks a least square fit $\Tilde{p}^\theta_i$ for the $\Tilde{p}_i$'s such that $\Tilde{p}^\theta_i\leq \Tilde{p}^\theta_j$ for $p_i^\theta\leq p_j^\theta$. It reduces to finding $\Tilde{p}^\theta_1,\ldots, \Tilde{p}^\theta_n$ that minimizes
\[
\sum_{i=1}^nw_i^{\textsc{iso}}(\Tilde{p}^\theta_i- \Tilde{p}_i)^2\text{, subject to }\Tilde{p}_i^\theta\leq \Tilde{p}^\theta_j\text{ whenever }p_i^\theta\leq p_j^\theta,
\]
where $\left(w_i^{\textsc{iso}}\right)_{i = 1, \ldots, n}$ denotes the weights associated to each pair $(p_i^\theta,\Tilde{p}_i)_{i=1,\ldots, n}$. The weights allows us to emphasize on specific data points in the same manner as the weights defined for the RMSE in \eqref{eq:sse_distance_pure}.  Since the $p_i^\theta$'s fall in a totally ordered space, a simple iterative procedure called the Pool Adjacent Violators Algorithm (PAVA) can be used.  The pseudo algorithm that describes the procedure is provided below:
\begin{enumerate}
\item Initialize the sequence of values to be the same as the data points $\Tilde{p}_i^\ast =\Tilde{p}_i$.

\item Iterate through the sequence and identify "violations," which occur when the current value is greater than the next value, that is 
$$
\Tilde{p}_i^\ast>\Tilde{p}_{i+1}^\ast\text{ for some }i=1,\ldots, n.
$$ 
When a violation is found, adjust the values in the associated segment of the sequence to be the average of the values,
$$
\Tilde{p}_i^\ast \leftarrow (\Tilde{p}_i^\ast + \Tilde{p}_{i+1}^\ast)/2,
$$ 
ensuring monotonicity.
\item Repeat Step 2 until no violations are left. 
\end{enumerate}
We use the \texttt{isoreg} function from $R$ to get the fitted values $\Tilde{p}^\theta_i,\text{ }i=1,\ldots, n$. To complete the isotonic regression task we shall find a function $f$ such that $f(p_i^\theta)=\Tilde{p}^\theta_i$. A common choice is a piece-wise constant function that interpolates the $\Tilde{p}^\theta_i$'s. An illustration is provided in \cref{ex:iso_fit} and we outline the advantages and limitations associated to this choice in \cref{rem:iso_reg}.
\begin{ex}\label{ex:iso_fit}
The isotonic fit of the data of \cref{ex:insurance_coverage} and \cref{ex:safety_loadings} is provided on \cref{fig:iso_plot}. 
\begin{figure}[!ht]
\centering
  \includegraphics[width=0.7\linewidth]{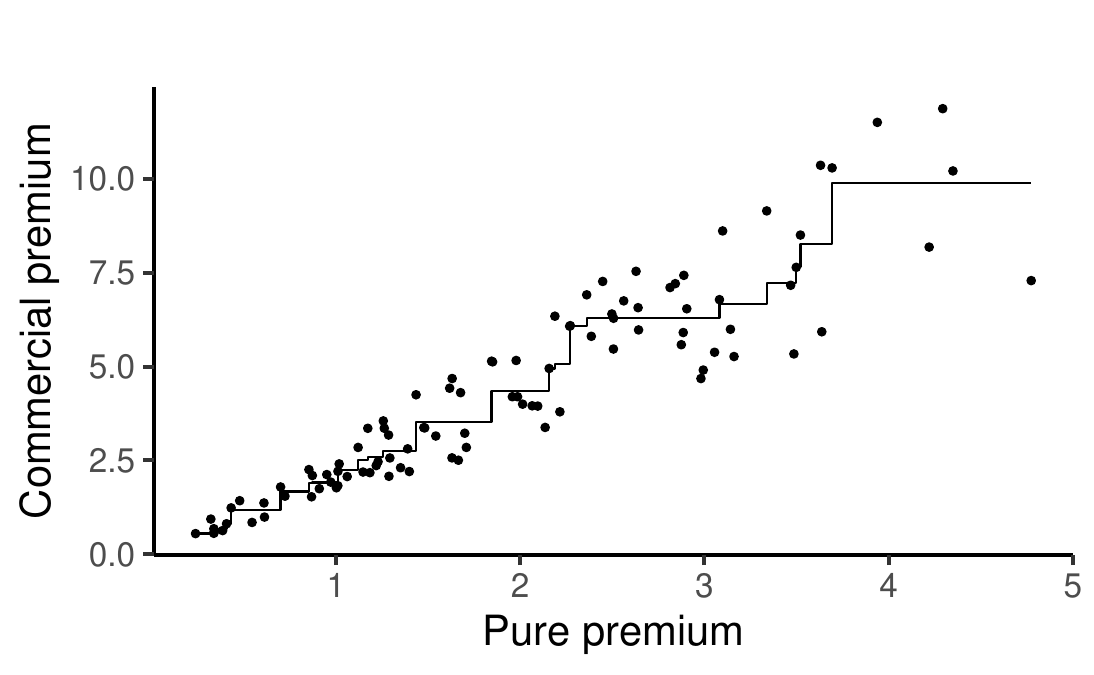}
  \caption{Isotonic link between the pure and commercial premiums.}
  \label{fig:iso_plot}
\end{figure}
\end{ex}
\begin{remark}\label{rem:iso_reg}
When looking at \cref{fig:iso_plot}, one may object that a simple linear regression model could do the job. This impression is partly due to the (noisy) linear link between pure and commercial premium in \cref{eq:cps}. Isotonic regression is a non-parametric approach, meaning it doesn't make strong assumptions about the underlying distribution or functional form of the relationship between variables. This can be advantageous when the true relationship is not well represented by a linear model. We briefly illustrate this fact in \cref{app:comparison_linear_isotonic} by looking at the residuals of the linear and isotonic regression and considering a non-linear link function between pure and commercial premium. Furthermore, we believe that ensuring $\Tilde{p}_i < \Tilde{p}_j$ when $p_i < p_j$ is desirable as greater pure premium should be associated to greater commercial premium as a rule of thumb. Isotonic regression aims at satisfying just this condition.  The main drawback of isotonic regression is that it can adhere too closely to the data points, risking overfitting. Note also that isotonic regression lacks the interpretability of a simpler, lower-dimensional parametric curve. 
\end{remark}

We now turn to the definition of a discrepancy measure to compare the model-generated and observed market commercial premiums.   Our starting point is the root mean square error (RMSE) defined as
\begin{equation}\label{eq:sse_distance}
\text{RMSE}\left[\Tilde{p}_{1:n}, f\left(p^\theta_{1:n}\right)\right] = \sqrt{\sum_{i=1}^nw^{\text{RMSE}}_i\left[\Tilde{p}_i - f\left(p_i^\theta\right)\right]^2},
\end{equation}
for candidate  risk parameter $\theta$ and the isotonic fit $f$ . We seek 
$$
\theta^\ast = \underset{\theta\in \Theta}{\argmin}\,
\text{RMSE}\left[\Tilde{p}_{1:n}, f\left(p^\theta_{1:n}\right)\right].
$$

The existence of such $\theta^\ast$ is guaranteed because $\theta\mapsto \text{RMSE}\left[\Tilde{p}_{1:n}, f\left(p^\theta_{1:n}\right)\right]$ only takes a finite number of values. Indeed, to each $\theta \in \Theta$ is associated a unique permutation $s^\theta\in S_n$, where $S_n$ denotes the set of all the permutations of $\{1,\ldots, n\}$, such that 
$$
p^\theta_{s^\theta(1)}\leq\ldots\leq p^\theta_{s^\theta(n)}.
$$
This permutation $s^\theta$ defines a unique isotonic fit $f$ based on 
$$
\Tilde{p}^\theta_{s^\theta(1)}\leq\ldots\leq \Tilde{p}^\theta_{s^\theta(n)},
$$
leading to a given RMSE value $\text{RMSE}\left[\Tilde{p}_{1:n}, f\left(p^\theta_{1:n}\right)\right]$. Concretely, for $\theta_1,\theta_2\in \Theta$, if it holds that $s^{\theta_1} = s^{\theta_2}$ then $\text{RMSE}\left[\Tilde{p}_{1:n}, f(p^{\theta_1}_{1:n})\right] = \text{RMSE}\left[\Tilde{p}_{1:n}, f(p^{\theta_2}_{1:n})\right]$. The application $\theta\mapsto s_n^\theta$ is surjective since $S_n^\Theta = \{s_n^\theta\text{ ; }\theta\in \Theta\}$  is finite.  The fact that $\Theta$ is a continuous space implies that $\theta^\ast$ cannot be unique. The application of isotonic regression offers a straightforward interpretation of our objective, as detailed in \cref{rem:iso_justification}.
\begin{remark}\label{rem:iso_justification}
At this stage, the optimization problem simplifies to identifying the parameter value $\theta$ corresponding to the most suitable permutation $s^\theta$ of the pure premium, which provides the best isotonic fit. We adhere to the guiding principle that a higher pure premium implies a higher commercial premium, as noted in \cref{rem:iso_reg}.
\end{remark}

Our problem is an ill-posed inverse problem. Ill-posedness is usually dealt with by adding a regularization to the objective function that one wants to minimize. The ratio of \(p / \Tilde{p}\) corresponds to what practitioners would call the expected Loss Ratio (\(\text{LR}\)). Our solution is based on targeting a given loss ratio.  The loss ratio is a standard measure to assess the profitability of insurance lines of business. An insurance company that enters a new market is likely to have insights on the loss ratio relative to this market, for example by having informal discussions with reinsurers, brokers or competitors. These feedbacks may translate into the definition of a lower and upper bound denoted by \(\text{LR}_{\text{low}}\) and \(\text{LR}_{\text{high}}\), respectively. We can then assume that the loss ratios \(\text{LR}_i = p_i / \Tilde{p}_i\), for \(i = 1, \ldots, n\), should fall in the range \([\text{LR}_{\text{low}}, \text{LR}_{\text{high}}]\), which we refer to as the loss ratio corridor. Assuming that \(\text{LR}_{\text{high}} < 1\), we ensure both constraints in \eqref{eq:constraint} by adding to our distance \eqref{eq:sse_distance} two regularization terms defined as
\begin{equation*}\label{eq:LR_bounds1}
\text{Reg}_{\text{low}}\left(\Tilde{p}_{1:n}, p_{1:n}^\theta\right) = \sqrt{\sum_{i=1}^nw_i^{\text{RMSE}}\left(\Tilde{p}_i - p_i^\theta \cdot \text{LR}_{\text{low}}^{-1}\right)_+^2}, 
\end{equation*}
and 
\begin{equation*}\label{eq:LR_bounds2}
\text{Reg}_{\text{high}}\left(\Tilde{p}_{1:n}, p_{1:n}^\theta\right) = \sqrt{\sum_{i=1}^nw_i^{\text{RMSE}}\left(p_i^\theta \cdot \text{LR}_{\text{high}}^{-1} - \Tilde{p}_i\right)_+^2},
\end{equation*}
where \((x)_+ = \max(x,0)\) denotes the positive part of \(x\). The distance we consider within \cref{pb:optimization_problem} is now given by 
\[
d\left[\Tilde{p}_{1:n}, f\left(p_{1:n}^\theta\right)\right] = \text{RMSE}\left[\Tilde{p}_{1:n}, f\left(p_{1:n}^\theta\right)\right] + \text{Reg}_{\text{low}}\left(\Tilde{p}_{1:n}, p_{1:n}^\theta\right) + \text{Reg}_{\text{high}}\left(\Tilde{p}_{1:n}, p_{1:n}^\theta\right).
\]
We illustrate the impact of adding the regularization terms in \cref{ex:before_after_reg}.

\begin{ex}\label{ex:before_after_reg}
We consider the commercial rates of \cref{ex:safety_loadings}. Recall that the commercial premiums are given by
\begin{equation}
\Tilde{p}_i = (1+\eta_i)p_i, \quad \text{for} \quad i=1, \ldots, n,
\end{equation}
where the pure premiums are those of \cref{ex:model_identifiability_pp} and
$$
\eta_i \sim \UnifDist([0.5, 2]), \quad \text{for} \quad i=1, \ldots, n.
$$
\cref{fig:contour_plot_w_wo_reg} displays the contour plot of the discrepancy between observed and model-generated commercial rates.
\begin{figure}[!ht]
  \begin{center}
    \subfloat[Distance function without regularization]{
      \includegraphics[width=0.5\textwidth]{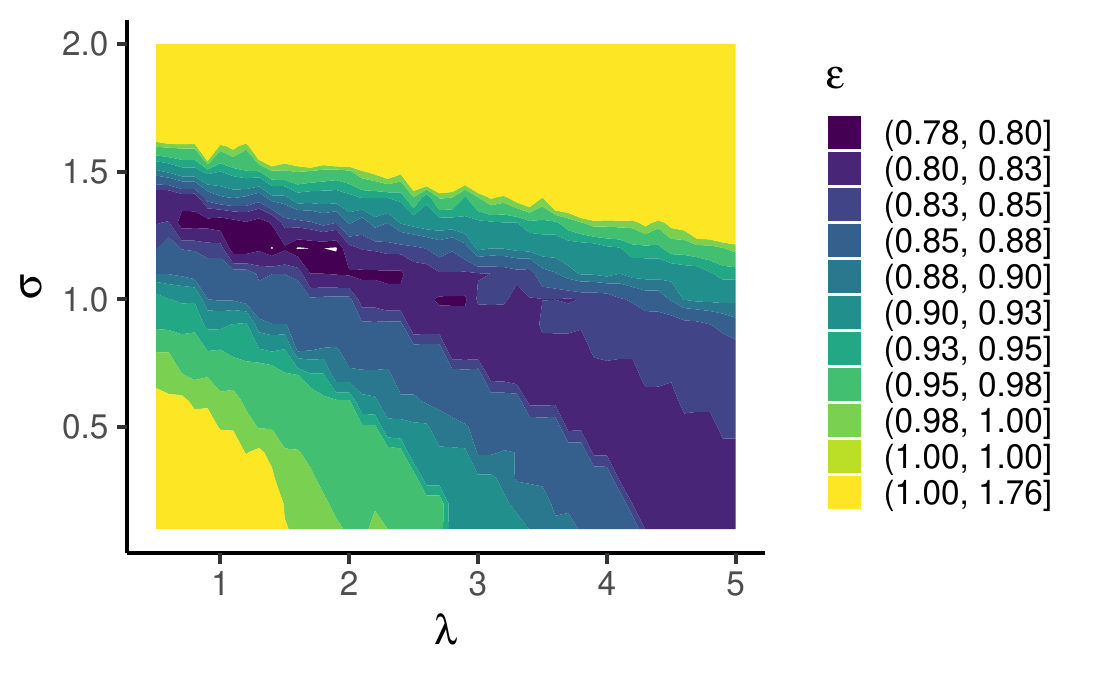}
      \label{sub:cp_contour_plot_wo_rg}
    }
    \subfloat[Distance function with regularization]{
      \includegraphics[width=0.5\textwidth]{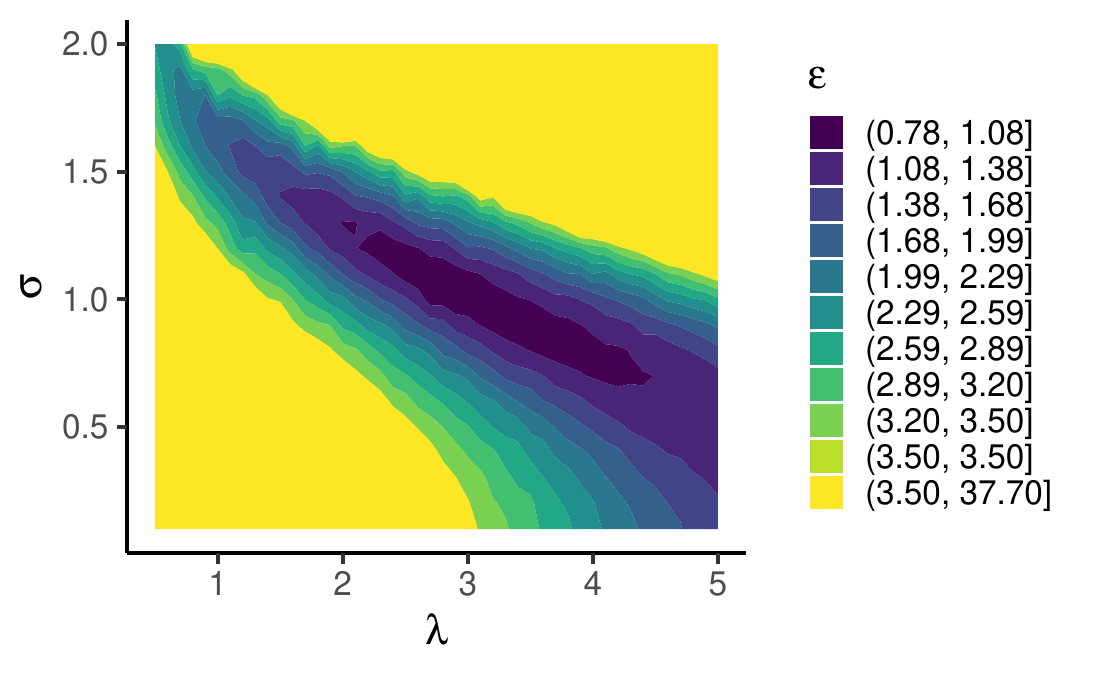}
      \label{sub:cp_contour_plot_w_rg}
    }
    \caption{Contour plot of \(\text{RMSE}\left[\Tilde{p}_{1:n}, f\left(p^\theta_{1:n}\right)\right]\) and \(d\left[\Tilde{p}_{1:n}, f\left(p_{1:n}^\theta\right)\right]\) for \(\mu = 0\) and \((\lambda, \sigma) \in [0, 5] \times [0, 2]\).}
    \label{fig:contour_plot_w_wo_reg}
  \end{center}
\end{figure}
When comparing \cref{sub:cp_contour_plot_wo_rg} and \cref{sub:cp_contour_plot_w_rg}, we note how beneficial including regularization terms is to identify the true parameter values.
\end{ex}

Regularization brings us closer to the scenario described in \cref{ssec:problem_1}, where the pure premium is known. This corresponds to the case where $\text{LR}_{\text{low}} = \text{LR}_{\text{high}} = 1$. Regularization enables us to exclude large portions of the parameter space associated with nonsensical pure premiums given the commercial premiums. However, careful consideration is required when setting the loss ratio corridor. A narrow loss ratio corridor results in a highly precise estimation, but this estimation may be biased if the loss ratio corridor is misspecified. 

The issue of identifiability persists, as multiple parameter values can still solve the optimization problem. This highlights the necessity of using particle-based optimization techniques, as described in \cref{ssec:algo}, to explore the parameter space. Such techniques return a set of admissible candidate parameters and are less sensitive to initialization. For a more detailed discussion of the identifiability issue, we refer the reader to the online supplementary material\footnote{\url{https://github.com/LaGauffre/market_based_insurance_ratemaking/blob/main/latex/supp_material.pdf}}.

\subsection{Population Monte Carlo Approximate Bayesian Computation algorithm}\label{ssec:algo}
Our solution alternates between proposing parameter values for the risk to compute the pure premiums and approximating the \(f_i\)'s using isotonic regression. We must accommodate the lack of tractable expressions for the pure premium
$$
p_i^\theta = \mathbb{E}_\theta\left[g_i(X)\right], \quad \text{for} \quad i = 1, \ldots, n.
$$
The use of numerical methods makes a grid search procedure prohibitive from a computing time point of view. It also prevents us from using gradient-based optimization procedures. In such cases, one can turn towards particle swarm optimization algorithms or genetic algorithms to search the parameter space. Since we have decided to take a Crude Monte Carlo estimator for the pure premiums, the accuracy depends on the number of replications \(R\) of \(X\) being used. We adopt a Bayesian strategy in order to reflect the uncertainty around the pure premium calculation onto the parameters' final estimates. Our algorithm is similar to Approximate Bayesian Computation algorithms and we simply refine the procedure laid out in the introduction to get an approximation of the posterior distribution \(\pi(\theta|\mathcal{D})\).

\noindent We start by setting a prior distribution \(\pi(\theta)\) over the parameter space that we sequentially improve through intermediate distributions characterized by a sequence of tolerance levels \((\epsilon_g)_{g\geq0}\) that decrease gradually as \(\infty = \epsilon_0 > \epsilon_1 > \epsilon_2 > \ldots > 0\). Each intermediate distribution (called a generation and denoted by $g$) is represented by a cloud of weighted particles\(\left(\theta_j, w_j^g\right)_{j=1, \ldots, J}\). We approximate each intermediate posterior distribution using a multivariate kernel density estimator (\kde) denoted by \(\pi_{\epsilon_g}(\theta|\mathcal{D})\). The parameters of the algorithm are the number of generations \(G\), the population size \(J\) (the number of particles in the cloud), and the number of Monte Carlo replications $R$ of $X$.

\noindent The algorithm is initialized by setting \(\epsilon_0 = \infty\) and \(\pi_{\epsilon_0}(\theta|\mathcal{D}) = \pi(\theta)\). For generation \(g \geq 1\), we hold an intermediate distribution \(\pi_{\epsilon_{g-1}}(\theta|\mathcal{D})\) from which we can sample particles $\theta^\ast \sim \pi_{\epsilon_{g-1}}(\theta|\mathcal{D})$. We compute the associated pure premium
$$
p_i^{\theta^{\ast}} = \mathbb{E}_{\theta^\ast}\left[g_i(X)\right], \quad \text{for} \quad i = 1, \ldots, n.
$$
The pure premiums are computed via Monte Carlo simulations. The accuracy depends on the number \(R\) of copies of \(X\) involved in the Monte Carlo estimations.
We then fit the isotonic regression model 
\begin{equation*}\label{eq:iso_reg}
\Tilde{p}_i = f\left(p_i^{\theta^{\ast}}\right) + e_i, \quad \text{for} \quad i = 1, \ldots, n,
\end{equation*}
where \(e_i\) is an error term that captures the mismatch between the true value of the pure premium and its empirical counterpart estimated by the competitor insurance company using its historical data and the company-specific loading function. We further compare the observed commercial premiums to the model-generated ones via the distance defined in \cref{ssec:optimization_problem_real} with 
\[
d\left[\Tilde{p}_{1:n}, f\left(p_{1:n}^{\theta^\ast}\right)\right] = \text{RMSE}\left[\Tilde{p}_{1:n}, f\left(p_{1:n}^{\theta^\ast}\right)\right] + \text{Reg}_1\left(\Tilde{p}_{1:n}, p_{1:n}^{\theta^\ast}\right) + \text{Reg}_2\left(\Tilde{p}_{1:n}, p_{1:n}^{\theta^\ast}\right).
\]
If the distance satisfies \(d\left[\Tilde{p}_{1:n}, f\left(p_{1:n}^{\theta^\ast}\right)\right] < \epsilon_{g-1}\), then we keep the associated particle \(\theta^\ast\). New particles are proposed until we reach \(J\) accepted particles denoted by \(\theta_1^g, \ldots, \theta^g_J\). We also store the distances \(d_1^g, \ldots, d_J^g\). We need to set the next tolerance threshold \(\epsilon_g\), which is used to calculate the particle weights
$$
w_j^g \propto \frac{\pi(\theta^g_j)}{\pi_{\epsilon_{g-1}}(\theta)}\mathbb{I}_{d_j^g < \epsilon_{g}}, \quad j = 1, \ldots, J.
$$
The tolerance threshold is chosen so as to maintain a specified effective sample size (\ess) of \(J/2\) as in \citet{DMDoJa12}. Following \citet{Kong1994}, the \ess is estimated by \(1/\sum_{j=1}^J(w_j^g)^2\). This weighted sample then allows us to update the intermediate distribution as 
$$
\pi_{\epsilon_g}(\theta|\mathcal{D}) = \sum_{j=1}^J w_j^g K_H(\theta - \theta^g_j),
$$
where \(K_H\) is a multivariate \kde with smoothing matrix \(H\). A common choice for the \kde is the multivariate Gaussian kernel with a smoothing matrix set to twice the empirical covariance matrix of the cloud of particles \(\{\theta_j^g, w_j^g\}\) as in \citet{beaumont2009adaptive}. The procedure is summarized in \cref{alg:PMC_abc}.

\begin{algorithm}%
  \caption{Population Monte Carlo Approximate Bayesian Computation}%
  \label{alg:PMC_abc}
  \begin{algorithmic}[1]
    \State \textbf{set} $\epsilon_0 = \infty$ and \(\pi_{\epsilon_0}(\bt \cond \mathcal{D}) = \pi(\bt)\)

    \For {\(g = 1 \to G\)}
    \For {\(j = 1 \to J\)}
    \Repeat
    \State \textbf{generate} \(\theta^\ast \sim \pi_{\epsilon_{g-1}}(\theta \cond \bx)\)
    \State \textbf{compute} $p^{\theta^\ast}_i = \mathbb{E}_{\theta^\ast}\left[g_i(X)\right]\text{, for }i = 1,\ldots, n$
    \State \textbf{fit} the isotonic regression model $\Tilde{p}_i= f(p_i^{\theta^\ast})+e_i \text{, for }i = 1,\ldots, n$
    \State \textbf{compute} $d\left[\Tilde{p}_{1:n}, f\left(p_{1:n}^{\theta^\ast}\right)\right] = \text{RMSE}\left[\Tilde{p}_{1:n}, f\left(p_{1:n}^{\theta^\ast}\right)\right] + \text{Reg}_1\left(\Tilde{p}_{1:n}, p_{1:n}^{\theta^\ast}\right) + \text{Reg}_2\left(\Tilde{p}_{1:n}, p_{1:n}^{\theta^\ast}\right)$.
    \Until $d\left[\Tilde{p}_{1:n}, f\left(p_{1:n}^{\theta^\ast}\right)\right] < \epsilon_{g}$
    \State \textbf{set} $\theta^g_j = \theta^\ast$ and $d_j^g = d^\ast$
    \EndFor
    \State \textbf{find} $\epsilon_g \le \epsilon_{g-1}$ so that \(\widehat{\text{\ess}} = \Bigl[\sum_{j=1}^J\left(w_j^g\right)^2\Bigr]^{-1} \approx J/2\), where
    \[w_j^g \propto\frac{\pi(\theta_j^g)}{\pi_{\epsilon_{g-1}}(\theta_j^g \cond \mathcal{D})} \ind_{d_j<\epsilon_g}, \quad j = 1,\ldots, J\]

    \State \textbf{compute} \(\pi_{\epsilon_g}(\theta \cond \mathcal{D}) = \sum_{j = 1}^J w_j^g K_{\bH}(  \theta-\theta_j^g  )\)
    \EndFor
  \end{algorithmic}
\end{algorithm}

\noindent The user must configure several aspects of the algorithm. The prior assumptions \(\pi(\theta)\) determine the parameter space that will be searched. The loss ratio corridor \([\text{LR}_{\text{low}}, \text{LR}_{\text{high}}]\) sets up the two regularization terms, ensuring that parameters associated with unreasonable pure premiums are excluded. The prior settings and loss ratio corridor can be guided by expert opinions. The population size \(J\) drives the quality of the posterior distributions approximations through the cloud of particles. A large \(J\) also enhances the chances of finding global optimums, as more particles improve the coverage of the parameter space. A greater number \(R\) of Monte Carlo simulations ensures the accuracy of the pure premium evaluation. Both \(R\) and \(J\) contribute to the stability of the algorithm's results over several runs. The number of generations \(G\) relates to the tolerance level \(\epsilon\), which in turn drives the narrowness of the posterior distribution output by the ABC algorithm. As one would expect, the computational time for the algorithm increases with higher values of \(R, J\) and \(G\). Therefore, the choice of suitable values for \(G\), \(J\), and \(R\) can be made in consideration of a predetermined computational time budget. A practical solution to set \(G\) is to stop the algorithm whenever the difference between two consecutive tolerance levels is lower than some threshold \(\Delta_\epsilon\)or if we reach a minimum tolerance level $\epsilon_{\min}$. Convergence results for ABC algorithms are readily available in the litterature as discussed in \cref{rem:abc_convergence}.
\begin{remark}\label{rem:abc_convergence}
The final output of our ABC algorithm is the following expression:
\begin{equation}\label{eq:abc_posterior}
\pi_{\epsilon_G}(\theta|\mathcal{D}) = \sum_{j=1}^J w_j^G K_H(\theta - \theta^G_j),
\end{equation}
which represents an approximate posterior distribution of $\theta$ given an \iid sample $x_{1:R}$ of $X$. The data $\mathcal{D}$ can be interpreted as a set of summary statistics calculated based on a sample $x_{1:R}$ of size $R$, which corresponds to the number of Monte Carlo replications used to calculate the pure premium. Several types of convergence can be studied:

\begin{enumerate}
    \item $R \to \infty$
    \item $\epsilon_G \to 0$
    \item $J \to \infty$
\end{enumerate}

Convergence (1) determines the accuracy of $d\left[\Tilde{p}_{1:n}, f\left(p_{1:n}^{\theta^\ast}\right)\right]$ when using simulation to evaluate the pure premium $p_{1:n}^\theta$. These are Monte Carlo estimators of $\mathbb{E}_\theta[g_i(X)]$ for $i = 1, \ldots, n$, and convergence occurs at a rate proportional to $1/\sqrt{R}$.

Regarding convergence (2), as $\epsilon_G$ decreases toward $0$, our ABC estimator converges to the distribution of $\theta$ conditional on $\mathcal{D}$. This is equivalent to the true posterior distribution only if $\mathcal{D}$ consists of sufficient statistics for the risk model $X$. \citet[Proposition 2]{rubio2013simple} justify the use of non-sufficient statistics, under certain conditions, to make inferences about $\theta$. Their results remain valid when using a kernel density estimator such as \eqref{eq:abc_posterior} to represent the posterior distribution. The bias of the ABC posterior has been estimated to be $\mathcal{O}(\epsilon_G^2)$ in \citet{Barber2015}.

Convergence (3) pertains to the empirical probability measure approaching the true probability measure. Central limit theorems for this convergence are discussed in \citet{DelMoral2006} and \citet{DelMoral2012}.

\end{remark} 
We illustrate the posterior distribution evolution along the algorithm iterations in \cref{ex:abc_posterior_convergence}.

\begin{ex}\label{ex:abc_posterior_convergence}
We follow up on \cref{ex:model_identifiability_pp} and \cref{ex:before_after_reg}. We aim to fit the model 
$$
X\sim\PoissonDist(\lambda)-\LognormalDist(\mu = 0, \sigma).
$$
The prior asumptions are as follows
$$
\lambda \sim \UnifDist([0,10]),\text{ and }\sigma \sim \UnifDist([0,5]).
$$
The algorithm parameters are set to
$$
J = 1000\text{, }R =  1000\text{, }\Delta_\epsilon = 0.1\text{, and }LR\in [0.3, 0.66].
$$
The algorithm halts at the $9^{\text{th}}$ generation reaching a tolerance level of $\epsilon = 0.87$. \cref{fig:abc_gen_post} shows the sequence intermediate posterior distributions for $\lambda$ and $\sigma$.
\begin{figure}[!ht]
  \begin{center}
  \subfloat[$p_{\epsilon_g}(\lambda|\mathcal{D})\text{,  }g = 1,\ldots, 9$.]{
      \includegraphics[width=0.4\textwidth]{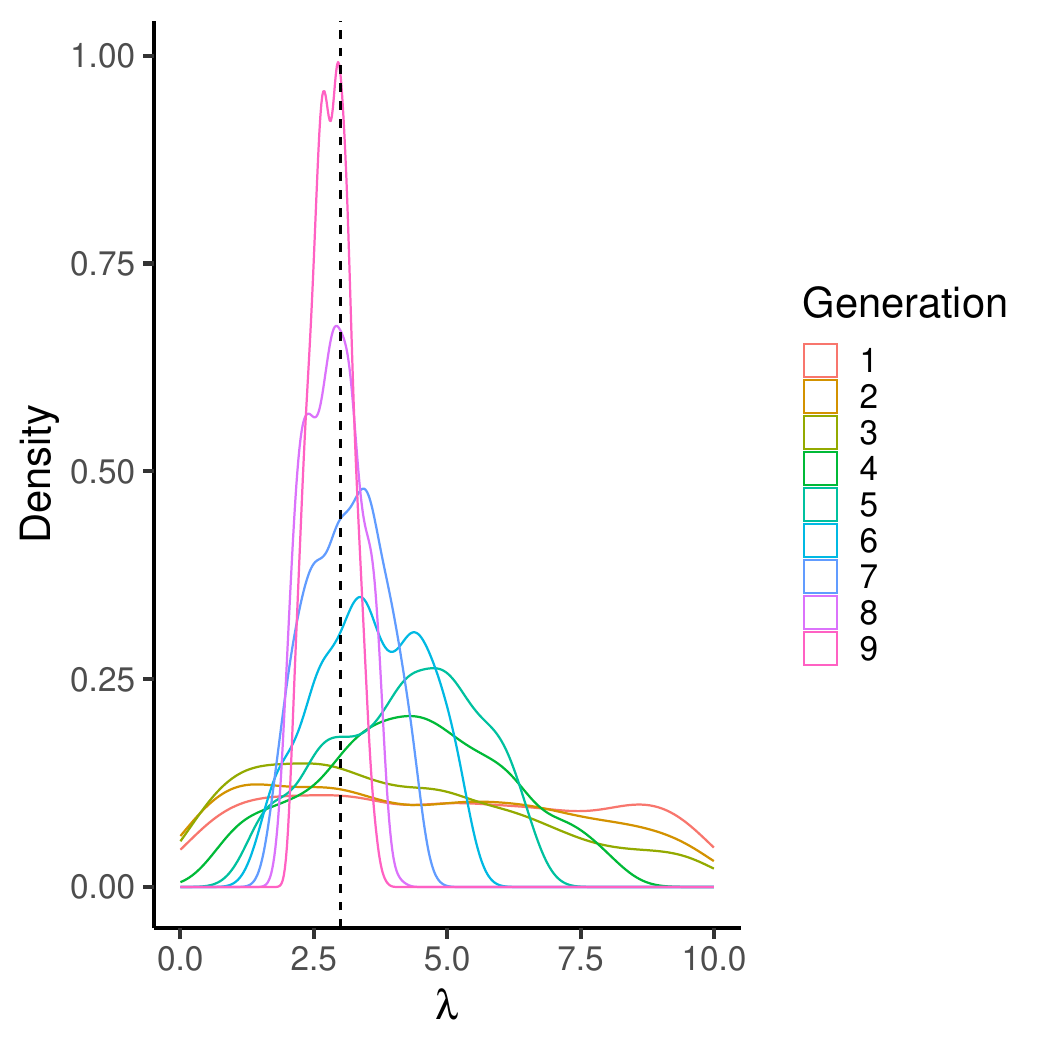}
      \label{sub:abc_gen_post_lambda}s
                         }
    \subfloat[$p_{\epsilon_g}(\lambda|\mathcal{D})\text{,  }g = 1,\ldots, 9$.]{
      \includegraphics[width=0.4\textwidth]{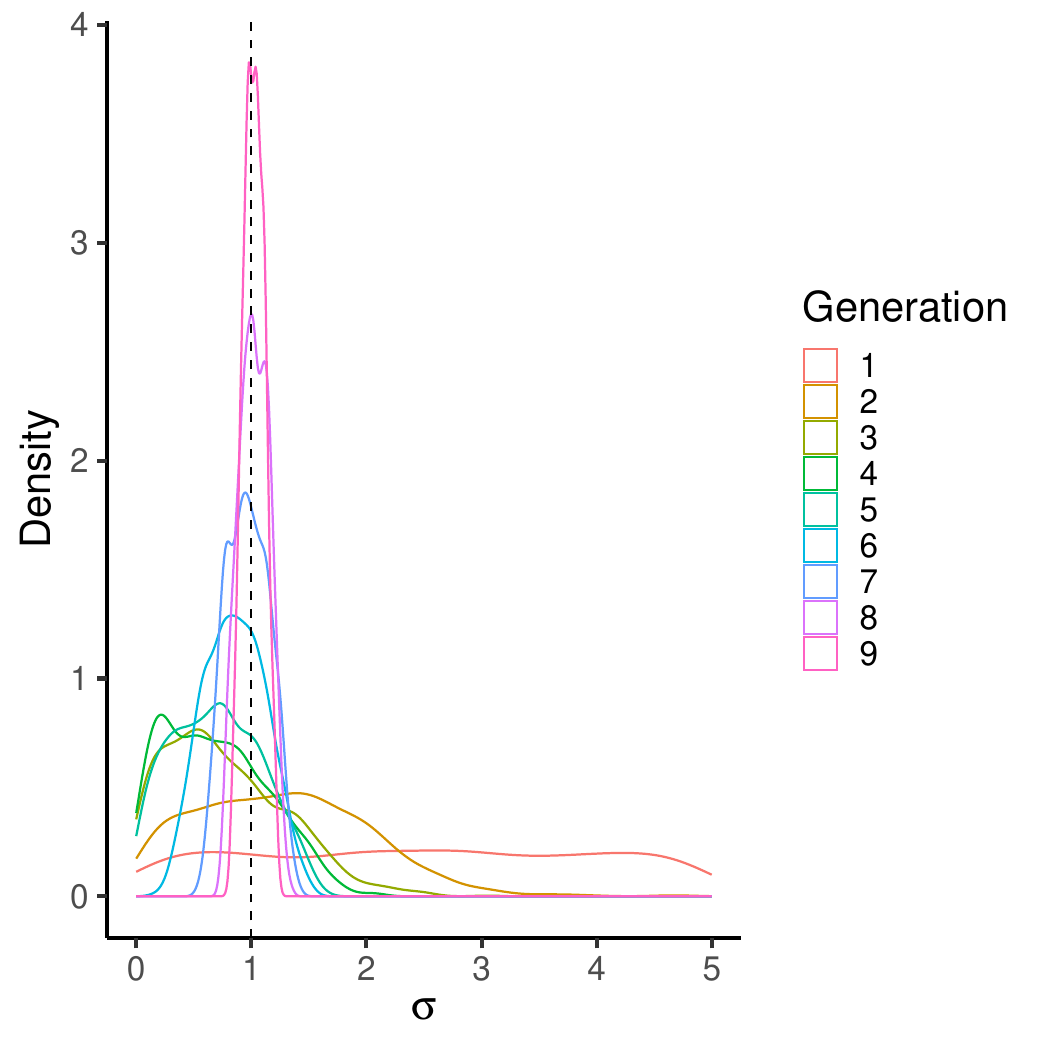}
      \label{sub:abc_gen_post_sigma}
                         }
\caption{Intermediate posterior distributions of $\lambda$ and $\sigma$.}
\label{fig:abc_gen_post}
  \end{center}
\end{figure}
\end{ex}


\noindent After the algorithm terminates, it is customary to focus on the last generations of particles for inference. Pointwise estimators are derived from this final set of particles. Two commonly used estimators include the Mean \textit{A Posteriori} (\map) obtained by averaging the particles in the last cloud and the Mode \textit{A Posteriori} (\mode), which is the mode of the empirical distribution within the final cloud of particles. The simulation study, conducted in the following section, is designed to investigate the convergence behavior and to compare the characteristics of the \map and \mode estimators.

\section{Methodology Assessment via Simulation}\label{sec:simulation}
\begin{figure}[!ht]
  \begin{center}
    \subfloat[\map and \bp for $\lambda$]{
      \includegraphics[width=0.4\textwidth]{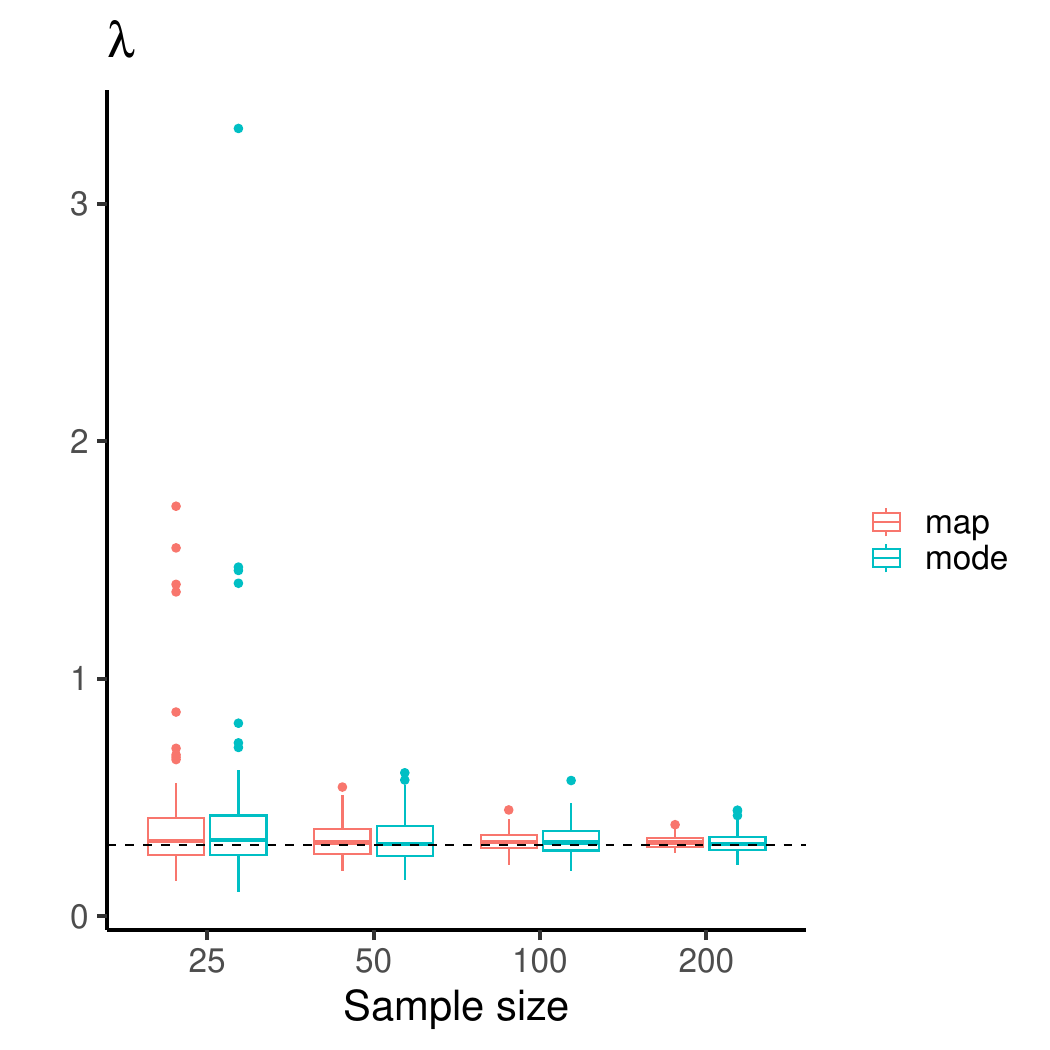}
      \label{sub:post_boxplot_simu_All_lambda}
                         }
    \subfloat[\map and \bp for $\mu$]{
      \includegraphics[width=0.4\textwidth]{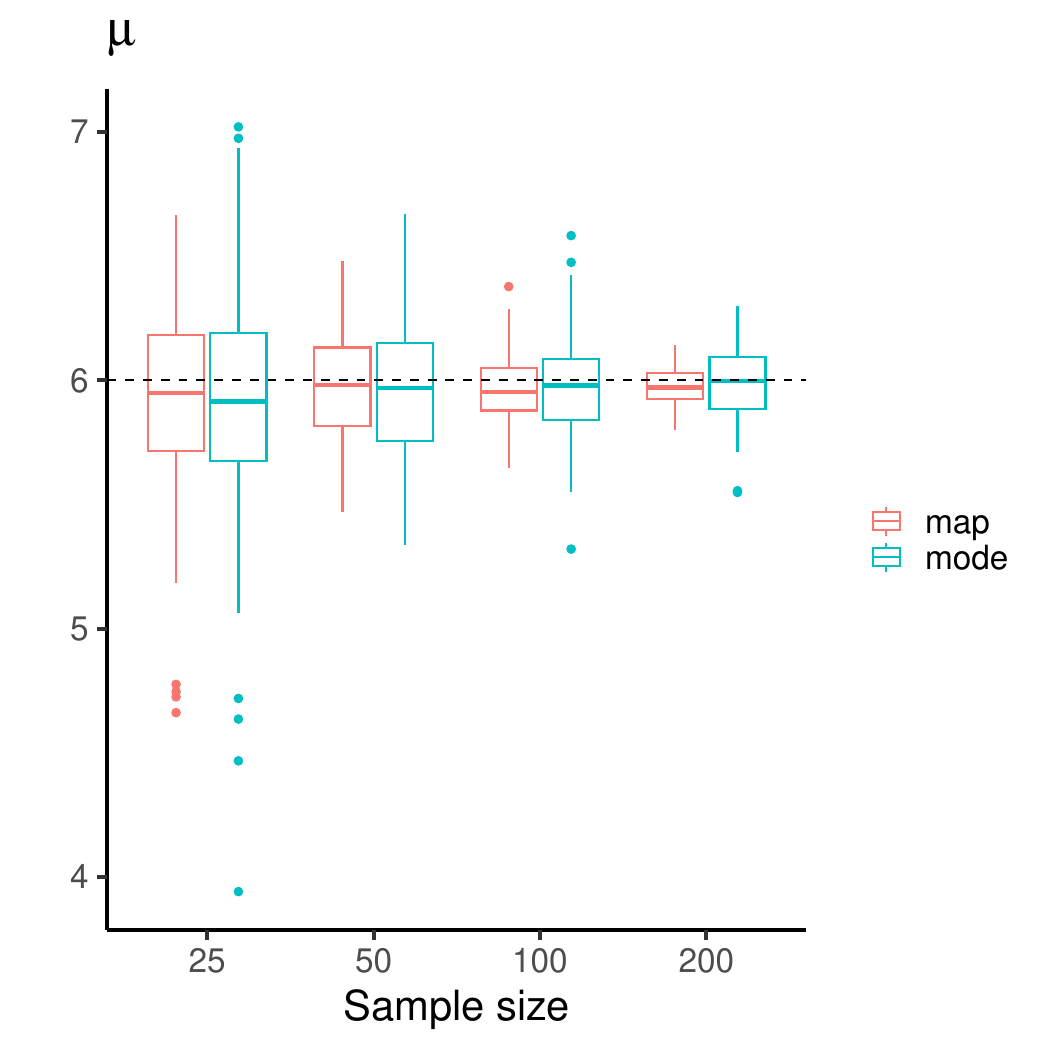}
      \label{sub:post_boxplot_simu_All_mu}
                         }
    \caption{\map and \mode estimators of the parameter of the model $\PoissonDist(\lambda=0.58)-\LognormalDist(\mu = 5.75, \sigma=1)$ based on synthetic market data of sizes $25, 50, 100,$ and $200$.}
    \label{fig:post_boxplot_simu_All}
  \end{center}
\end{figure}

In this section, we embark on an empirical exploration, seeking to understand how the posterior distribution of the parameters behaves as the sample size $n$ increases. This experimentation has been designed to resemble as much as possible the real data situation considered in \cref{ssec:one_risk_class_several_models}. We consider the risk, within a particular risk class, to be distributed as the random variable
$$
X = \sum_{k=1}^{N}U_k,
$$
where 
\begin{equation}\label{eq:claim_frequency_dist}
N\sim\PoissonDist(\lambda = 0.3),
\end{equation}
and
\begin{equation}\label{eq:claim_amount_dist}
U_k\sim \LognormalDist(\mu= 6, \sigma = 1),\text{ }k = 1,\ldots, N.
\end{equation}
The $U_i$'s are \iid and independent from $N$. We suppose that we know the variance parameter $\sigma$ and we try to draw inference on $\lambda$ and $\mu$. The parameter values of the claim frequency and severity in \eqref{eq:claim_frequency_dist} and \eqref{eq:claim_amount_dist} respectively are those infered in \cref{ssec:one_risk_class_several_models} for the $\PoissonDist-\LognormalDist$ model using the \mode estimator. The prior distributions are set to independent uniforms for $\lambda$ and $\mu$ as
$$
\lambda \sim \UnifDist([0, 10]),\text{ and }\mu\sim \UnifDist([-10, 10]).
$$
We generate artificial synthetic commercial premiums for this case study according to
$$
\Tilde{p}_i = (1+\eta_i)\mathbb{E}[g_i(X)] = (1+\eta_i)\mathbb{E}\{\text{min}[\text{max}(r_i\cdot X - d_i, 0), l_i]\},\text{ }i = 1,\ldots, n,
$$
where the premium parameters $r$, $d$ and $l$ are sampled from that of the real data considered in \cref{sec:petApplication}, so that the simulated data is as close as possible to the real data. The $\eta_i$'s are \iid from $\eta_i\sim\UnifDist([1.43, 2.5])$, which corresponds to loss ratios between $40\%$ and $70\%$. We further set $\text{LR}_\text{low} = 40\%$ and $\text{LR}_\text{high} = 70\%$. We consider sample of sizes $25,50, 100,$ and  $200$. We configure the algorithm with a population size of $J = 1,000$ and use $R = 2,000$ Monte Carlo replications. To ensure the algorithm's efficiency, we set a stopping threshold, requiring that the difference between two consecutive tolerance levels is smaller than $\Delta_\epsilon = 1$ for the algorithm to halt. These settings are kept for the analysis of real-world data, as they strike a balanced compromise between accuracy and computing time. We generate $100$ samples of fake data and apply our procedure. Our goal is to compare the result obtained using our two pointwise estimators: the mean \textit{a posteriori} \map and the mode \textit{a posteriori} \mode. The estimators of the parameters $\lambda$ and $\mu$ are given on \cref{fig:post_boxplot_simu_All}.

Both of the point-wise estimators seem to converge toward the parameter values that generated the data. The \map exhibits a better behavior than the \mode as its variability decreases in a notable way as the sample size increases.

\begin{figure}[!ht]
  \begin{center}

    \subfloat[Posterior distribution of $\mathbb{E}(U)$]{
      \includegraphics[width=0.3\textwidth]{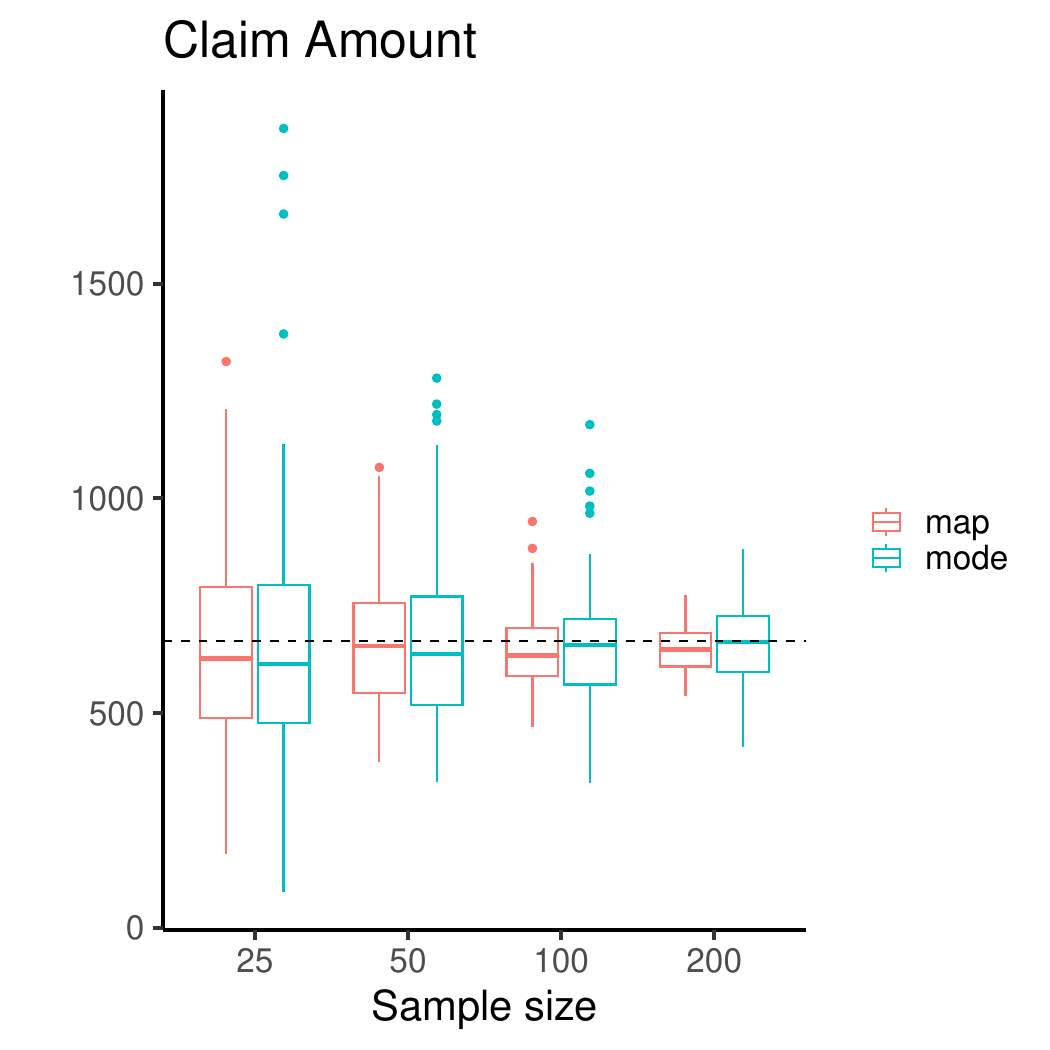}
      \label{sub:post_mu_sim}
                         }
    \subfloat[Posterior distribution of $\mathbb{E}(N)$]{
      \includegraphics[width=0.3\textwidth]{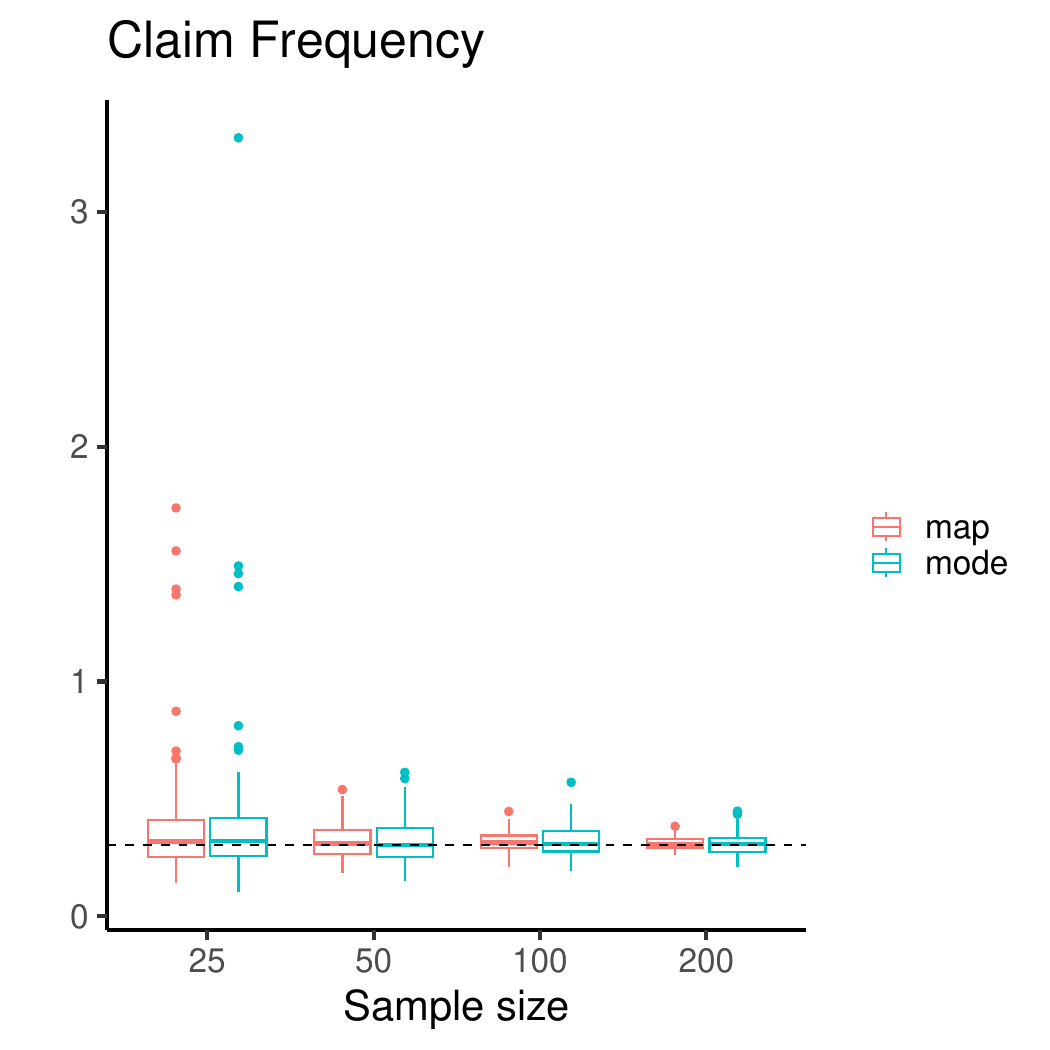}
      \label{sub:post_boxplot_simu_All_Claim_Frequency}
                         }
                         \hskip1em
    \subfloat[Posterior distribution of $\mathbb{P}(N=0)$]{
      \includegraphics[width=0.3\textwidth]{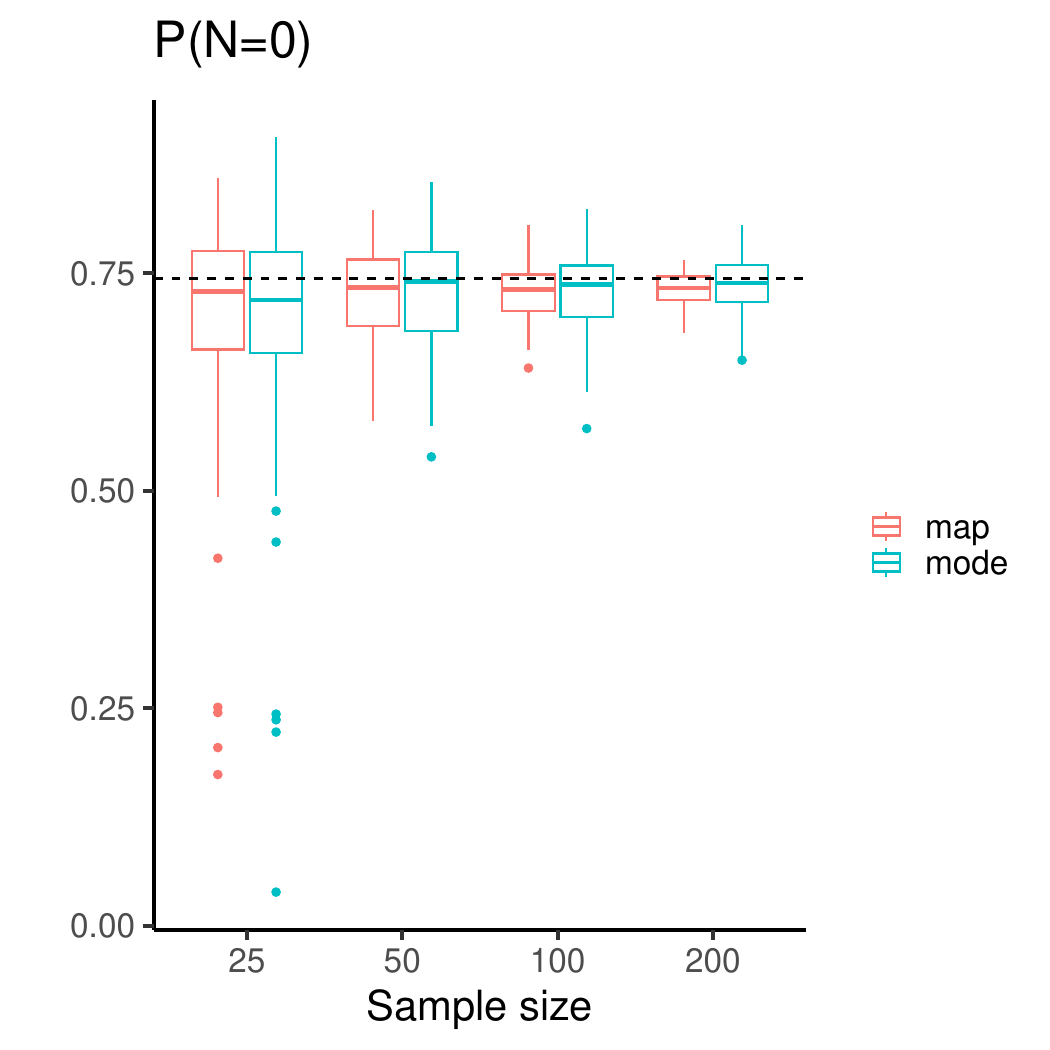}
      \label{sub:post_boxplot_simu_All_Prob_0}
                         }
    \subfloat[Posterior distribution of $\mathbb{E}(X)$]{
      \includegraphics[width=0.3\textwidth]{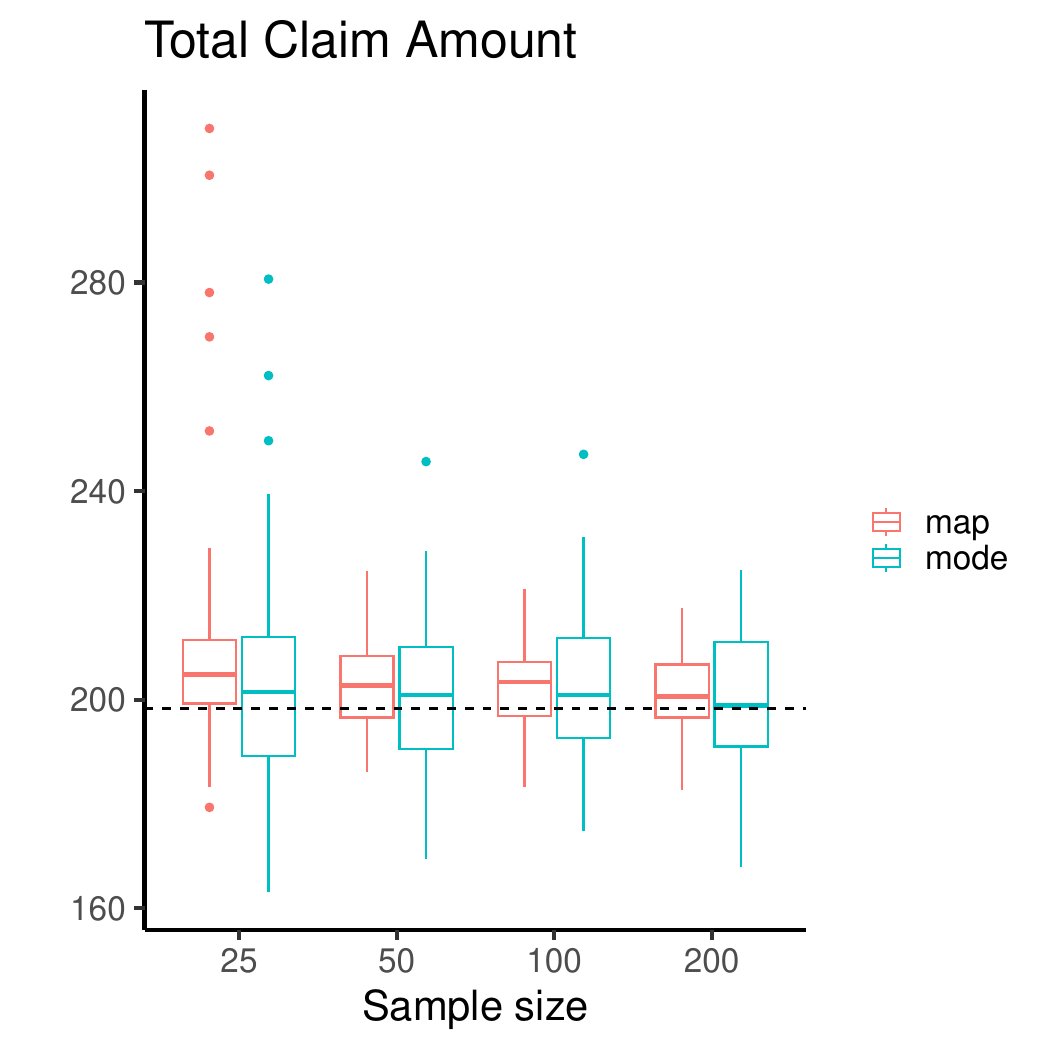}
      \label{sub:post_boxplot_simu_All_Total_Claim_Amount}
                         }   
                         \hskip1em                                  
    \subfloat[Posterior distribution of $\text{LR}$]{
      \includegraphics[width=0.3\textwidth]{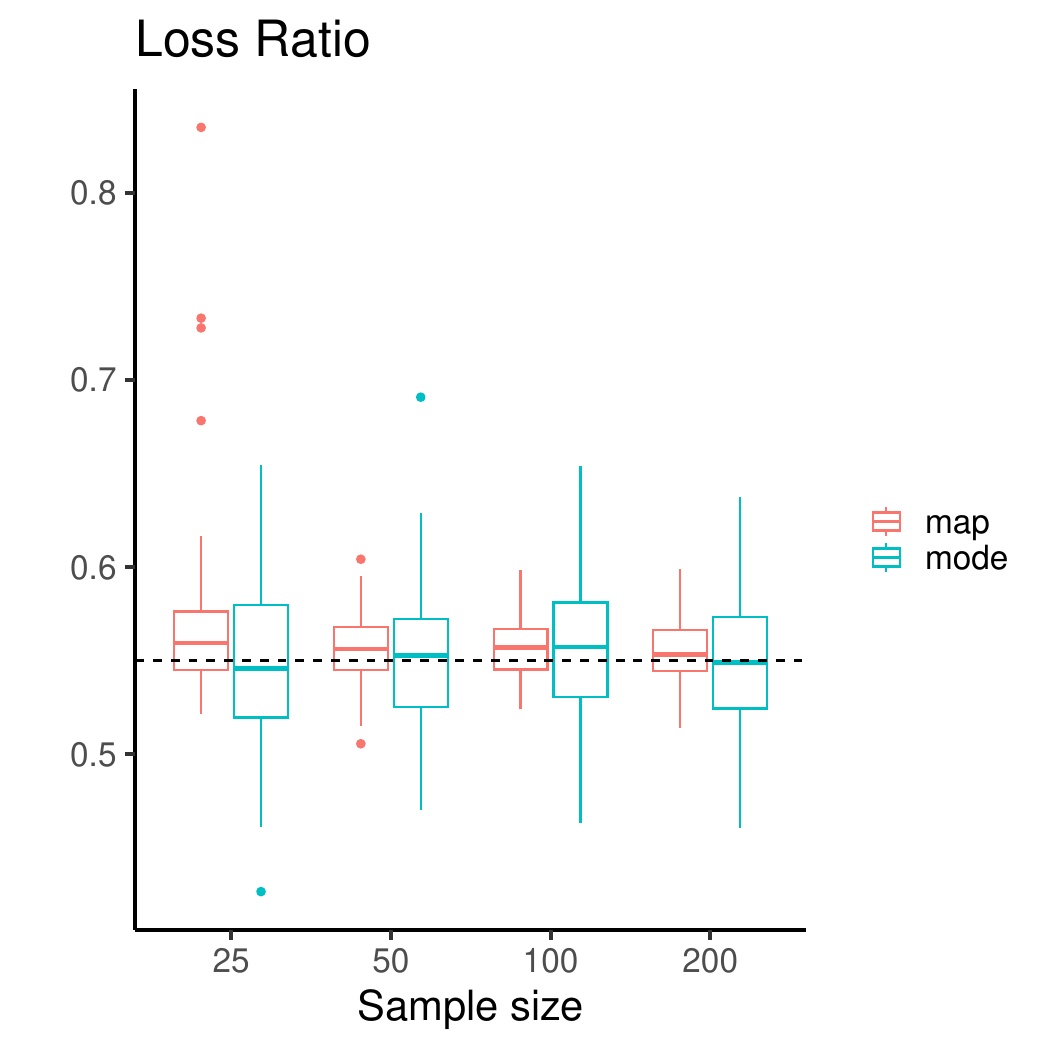}
      \label{sub:post_boxplot_simu_All_LR}
                         }         

    \caption{\map and \mode estimator of the features of the $\PoissonDist(\lambda=0.3)-\LognormalDist(\mu = 6, \sigma=1)$ loss model based on synthetic market data of sizes $25, 50, 100,$ and $200$.}
    \label{fig:post_boxplot_simu_All_key}
  \end{center}
\end{figure}

In Figure \ref{fig:post_boxplot_simu_All_key}, we present a comparison of key metrics, including the average claim amount, the average claim frequency, the probability of no reported claims, the average total claim amount, and the average loss ratio , defined as 
$$
\overline{\text{LR}} = \frac{1}{n}\sum_{i = 1}^{n}\frac{p_i}{\Tilde{p}_i}.
$$

Both estimation methods yield satisfactory results in recovering the characteristics of the loss distribution but the use of the \map yields more reliable estimations. 

\section{Application to the pet insurance market}%
\label{sec:petApplication} 

\subsection{Evolution and growth of the pet insurance market}
Pet insurance is a product designed to cover the costs of veterinary care for pets. It operates on a similar principle to human health insurance, providing a way for pet owners to manage the financial risks associated with unexpected medical expenses for their animals. Usually the expenses are covered in case of an accident or a disease. Pet owners can choose from different policy options based on their budget and coverage needs. Policies may vary in terms of deductibles ($d$), coverage limits ($l$), and coverage rates ($r$). The cost of premiums can depend on various factors, including the pet's age, breed, health condition, and the level of coverage selected.

The pet insurance market has been witnessing significant growth globally, driven by increasing pet ownership (especially with so-called pandemic pets, i.e. animals adopted during 2020 lockdowns), rising veterinary costs and the changing role that a pet plays in a families social structure. This latter factor is also influenced by changing societal views and the increased awareness of the importance of health and welfare of pets, which in turn comes with increased consideration of regular veterinary health checks . In order to offset the cost associated with such expenditures, there has begun to be a broader interest in households purchasing pet insurance. 

To date, the adoption and acceptance of pet insurance still varies significantly across regions of the world. Nordic countries, such as Sweden, have historically had a very high penetration rate with around $70\%$ of pets insured. Some Anglo-Saxon countries (UK and Germany mostly) have seen significant growth in the pet insurance market during the last decades, leading to $30\%$ of penetration rate. Other developed countries, like France, have significantly lower market sizes, with less than $10\%$ of pets that are insured, which suggests high growth potential. The market place for pet insurance in the USA is currently also experiencing sustained growth. According to MarketWatch nationwide survey\footnote{\url{https://www.marketwatch.com/guides/pet-insurance/pet-insurance-facts-and-statistics/}}, about $44.6\%$, of pet owners stated they currently have pet insurance. The North American Pet Health Insurance Association (NAPHIA) conducted a survey for its 2022 \textit{State of the Industry Report} and found that over $4.41$ million pets were insured in North America in 2021, up from $3.45$ million in 2020. The report also revealed that pet insurance premiums totaled $\$2.84$ billion in 2021, marking a $30.5\%$ increase from the previous year.



This growth continues to spur increases in the capital investments associated with such an insurance line of business:
\begin{itemize}
    \item in Sweden, Lassie has raised 11m euros in 2022  and 23m euros in 2023 ;
    \item in the UK, ManyPets has raised \$350m at a valuation higher than \$2bn in 2021;
    \item in France, Dalma has raised 15m euros in 2022.
    \item  JAB Holding Company has invested around 2 billion dollars in 2021 to create the Pinnacle Pet Group and the Independence Pet Holdings. Their purpose is to become the pet insurance leaders respectively in Europe and North America through multiple acquisitions of historic players. 
\end{itemize}

Hence, the pet insurance market is becoming more competitive with an increasing number of insurance companies or brokers offering pet insurance policies. To capture new market share as a new agent, it is essential to offer differentiated products, such as coverage combinations with no deductibles and higher limits.

\subsection{Data description}
During the week of the $18\textsuperscript{th}$ of May 2024, we have collected $1,080$ quotes from $5$ insurance companies. 
Each row of our datasets corresponds to a yearly premium collected from some insurance company website associated to a specific insurance coverage and a specific dog. We therefore find the coverage parameters which are the coverage rate $r$, the deductible $d$ and the limit $l$. Recall that the compensation for an annual expense of amount \( X \) is calculated as \(\min\left[\max(r \cdot X - d, 0), l\right]\).
We also have the rating factors which reduces for pet insurance in France to specie, breed, age and gender. \cref{tab:data_dictionary} provides a list of the variables in the datasets.
\begin{table}[!ht]
\footnotesize
\centering
\begin{tabular}{llll}
  \hline
Variable & Type &Description &Example \\ 
  \hline

  specie & character &Specie of the pet&dog \\ 
  breed & character &Breed of the pet& australian sheperd \\ 
  gender & character &Gender of the pet& female \\ 
  insurance\_carrier & character& identification number of the insurance company & 1 \\ 
  age & numeric &Age of the pet (in years) & 4 years \\ 
  
  r & numeric &Value of the coverage rate& 0.6 \\ 
  l & numeric &Value of the limit of the insurance coverage& 1100 \\ 
  d & numeric &Value of the deductible of the insurance coverage& 0 \\ 
  x & numeric &Yearly commercial premium& 234.33 \\ 
   \hline
\end{tabular}
\caption{List of the variables of our datasets}
\label{tab:data_dictionary}
\end{table}

The first five rows are given in \cref{tab:5_rows}.

\begin{table}[!ht]
\footnotesize
\centering
\begin{tabular}{lllrrrrlr}
  \hline
specie & breed & gender & insurance\_carrier & r & l & d & age & x \\ 
  \hline
dog & australian sheperd & female & 1 & 0.60 & 1100.00 & 0.00 & 2 years & 221.34  \\ 
  dog & australian sheperd & female & 1 & 0.70 & 1500.00 & 20.00 & 2 years & 290.62  \\ 
  dog & australian sheperd & female & 1 & 0.80 & 1800.00 & 30.00 & 2 years & 361.53  \\ 
  dog & australian sheperd & female & 1 & 1.00 & 2500.00 & 75.00 & 2 years & 739.27  \\ 
  dog & australian sheperd & female & 1 & 0.90 & 2200.00 & 50.00 & 2 years & 594.28  \\ 
   \hline
\end{tabular}
\caption{First five rows of our datasets}
\label{tab:5_rows}
\end{table}

We have collected rates associated to $12$ risk classes given in \cref{tab:risk_classes}.

\begin{figure}[!ht]
  \begin{center}
     \subfloat[Rates of coverage of the insurance policies]{
      \includegraphics[width=0.3\textwidth]{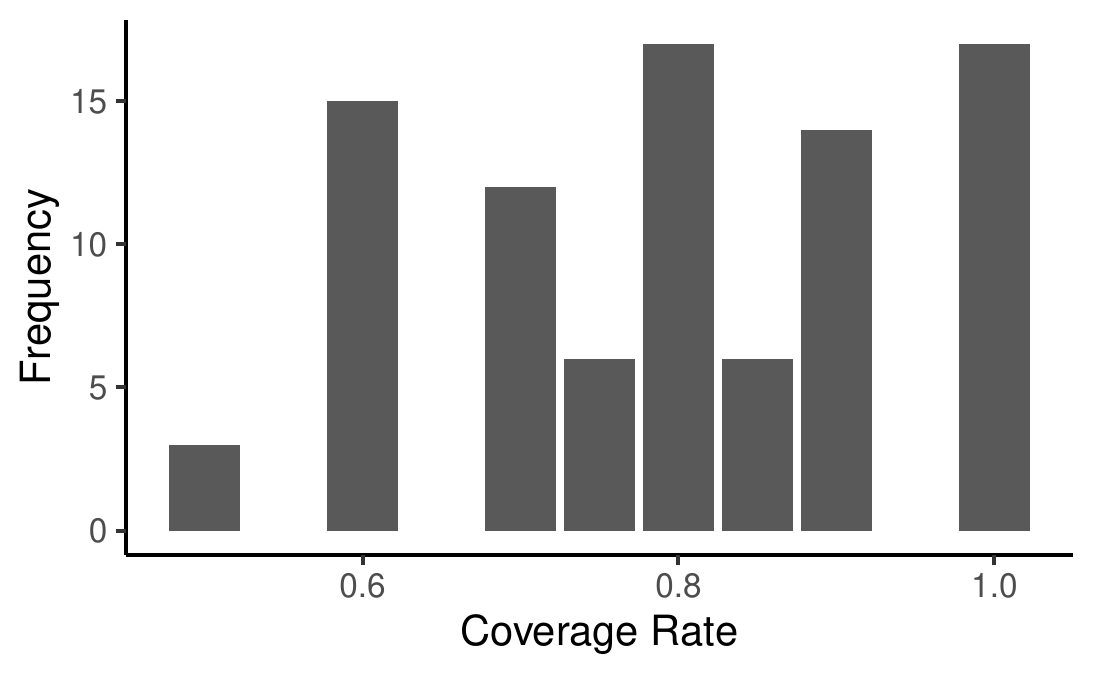}
      \label{sub:b_plot_r}
                         }
    \subfloat[Deductible of the insurance policies]{
      \includegraphics[width=0.3\textwidth]{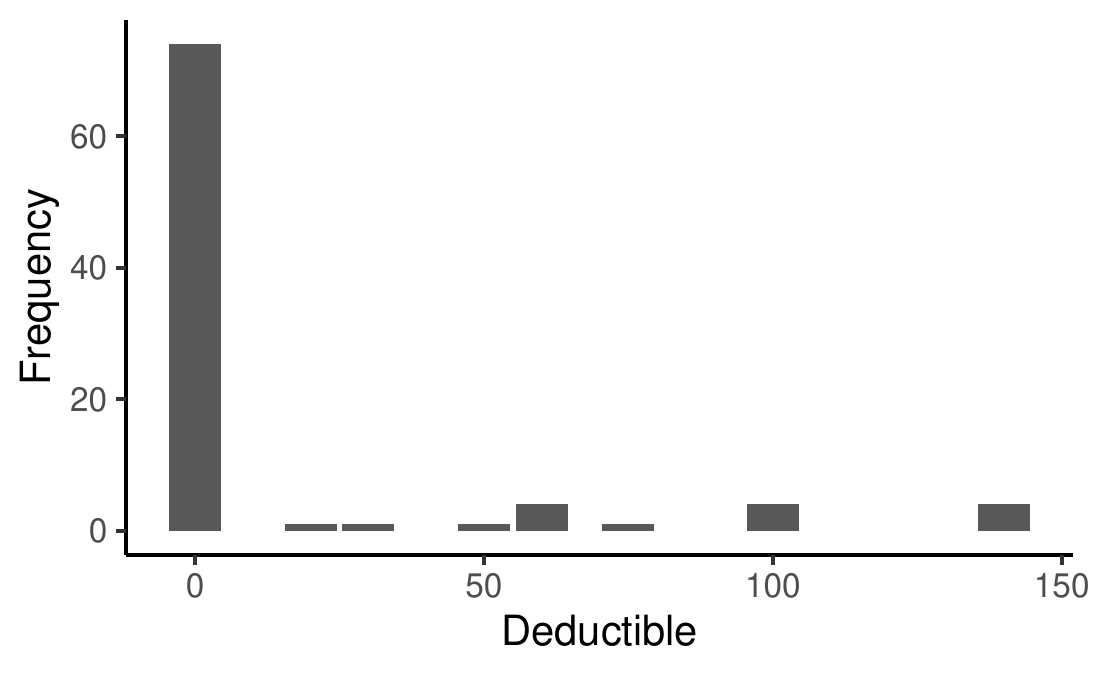}
      \label{sub:b_plot_d}
                         }
                         \hskip1em
    \subfloat[Limits of the insurance policies]{
      \includegraphics[width=0.3\textwidth]{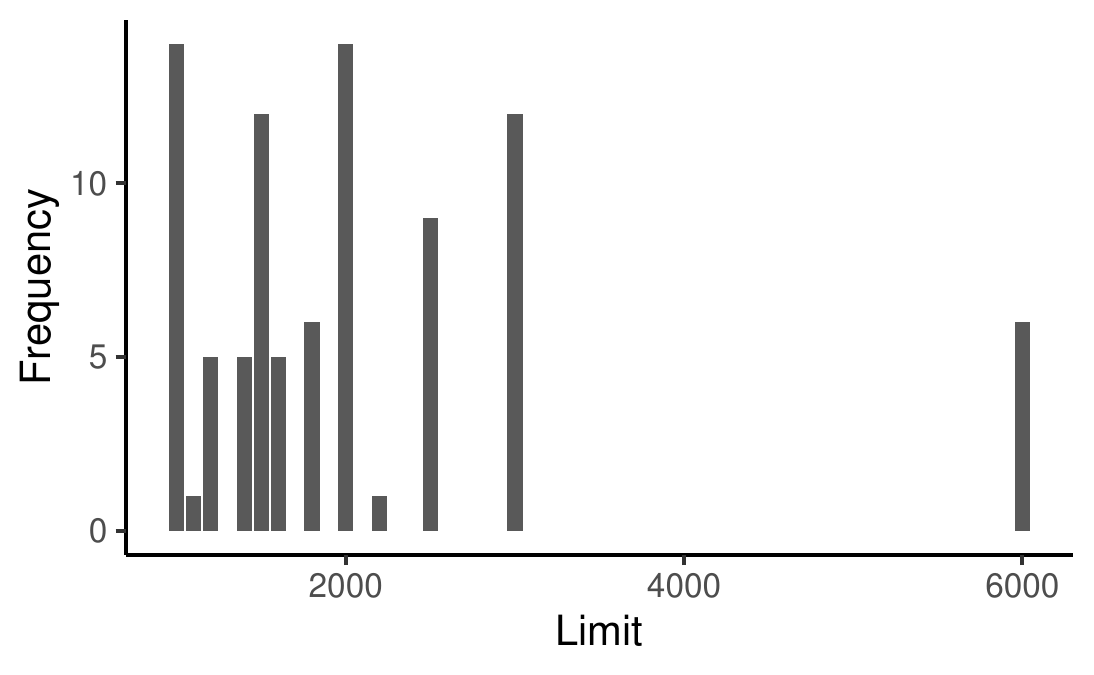}
      \label{sub:b_plot_l}
                         }


    \caption{Overview of the insurance coverages offered by the five insurance companies operating in the French market under study.}
    \label{fig:insurance_coverage}
  \end{center}
\end{figure}

\begin{table}[!ht]
\footnotesize
\centering
\begin{tabular}{l|llll}
  \hline
Risk class $\#$&specie & breed & gender & age \\ 
  \hline
1 & dog & australian sheperd & female & 4 months \\ 
  2 & dog & australian sheperd & female & 2 years \\ 
  3 & dog & australian sheperd & female & 4 years \\ 
  4 & dog & french bulldog & female & 4 months \\ 
  5 & dog & french bulldog & female & 2 years \\ 
  6 & dog & french bulldog & female & 4 years \\ 
  7 & dog & german sheperd & female & 4 months \\ 
  8 & dog & german sheperd & female & 2 years \\ 
  9 & dog & german sheperd & female & 4 years \\ 
  10 & dog & golden-retriever & female & 4 months \\ 
  11 & dog & golden-retriever & female & 2 years \\ 
  12 & dog & golden-retriever & female & 4 years \\ 
   \hline
\end{tabular}
\caption{The $12$ risk classes under study}
\label{tab:risk_classes}
\end{table}
It means that we have $90$ quotes to study each risk class. Figure \ref{fig:insurance_coverage} provides a visual overview of the range of insurance coverage options available in the pet insurance market. 

\cref{rem:small_print} briefly discusses data quality.

\begin{remark}\label{rem:small_print}
This analysis assumes that all insurance companies cover the same type of risk. While this assumption may not always hold in general insurance markets, it was valid for the French pet insurance market in May 2024, when the data for this study was collected. At that time, most insurers offered very similar coverage, with nearly identical exclusion clauses in their contracts. Only one outlier, a company that imposed sub-limits on specific procedures, was identified and excluded from this study. This homogeneity reflects the nascent stage of the French pet insurance market, where insurers tend to adhere to similar guidelines, and product innovation and differentiation have yet to emerge.
\end{remark}

We conduct two separate studies. In \cref{ssec:one_risk_class_several_models}, we focus on  a specific risk class associated to a female, 4 years old, australian sheperd. Several claim models are compared. One is selected to look into the pricing strategies of the actors. In \cref{ssec:several_risk_classes_one_model}, we look into the quotes of various risk classes that we investigate using a single model.

\subsection{Analysis of one risk class using several models}\label{ssec:one_risk_class_several_models}
\begin{figure}[!ht]
  \begin{center}

     \subfloat[ $\PoissonDist(\lambda) - \LognormalDist(\mu, \sigma = 1)$]{
      \includegraphics[width=0.3\textwidth]{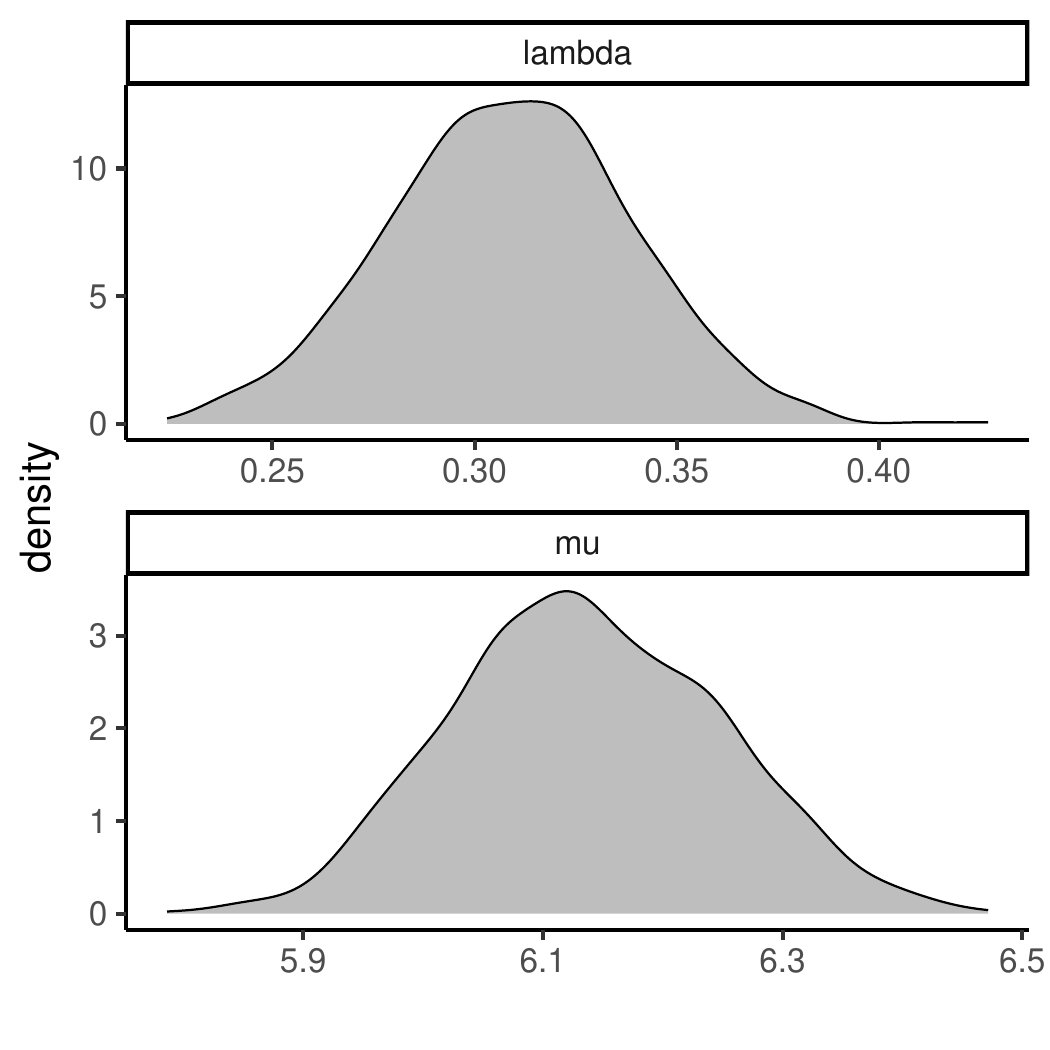}
      \label{sub:post_plot_Pois-Lognormalsigma1)}
                         }
  \subfloat[ $\PoissonDist(\lambda) - \LognormalDist(\mu=0, \sigma)$]{
      \includegraphics[width=0.3\textwidth]{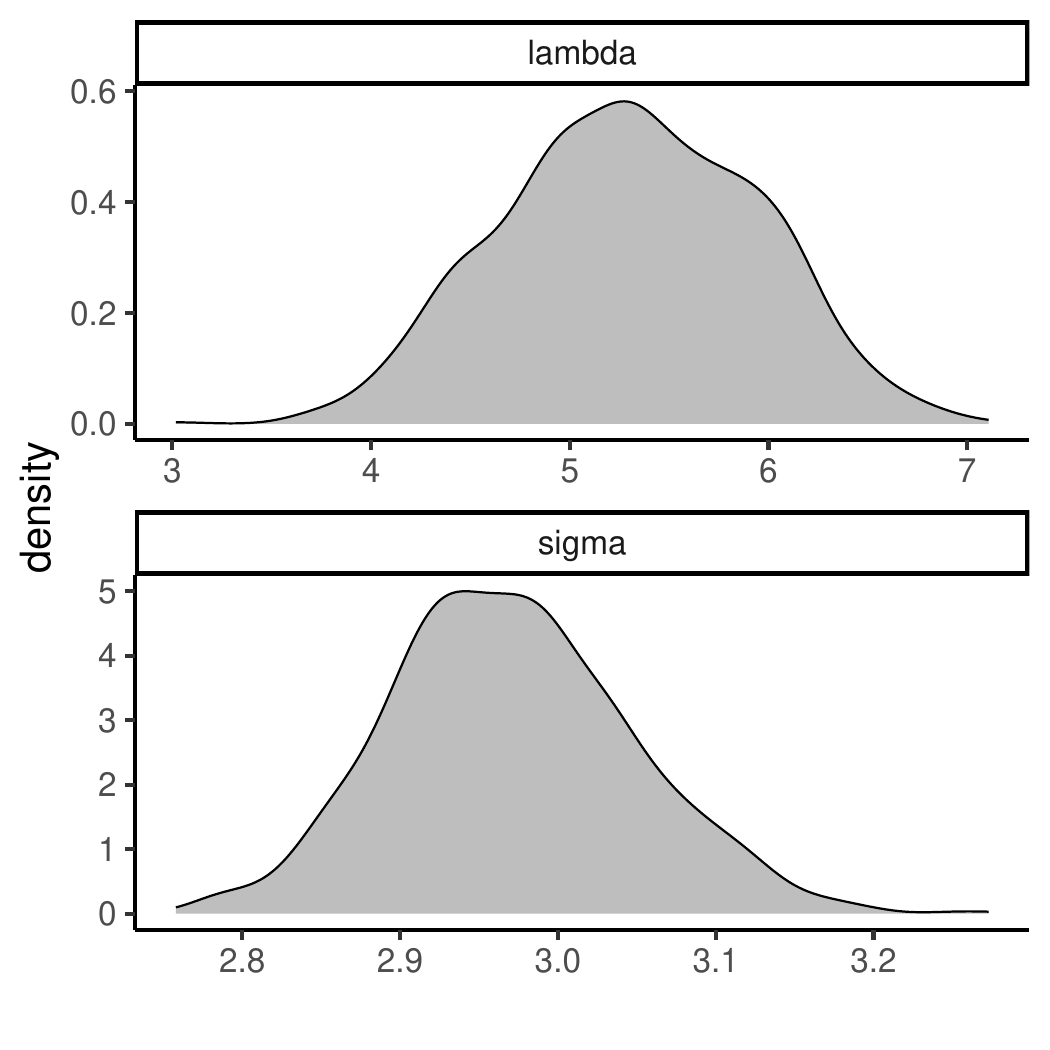}
      \label{sub:post_plot_Pois-Lognormalmu0)}
                         }
  \subfloat[$\PoissonDist(\lambda) - \GammaDist(\alpha, \beta=1)$]{
      \includegraphics[width=0.3\textwidth]{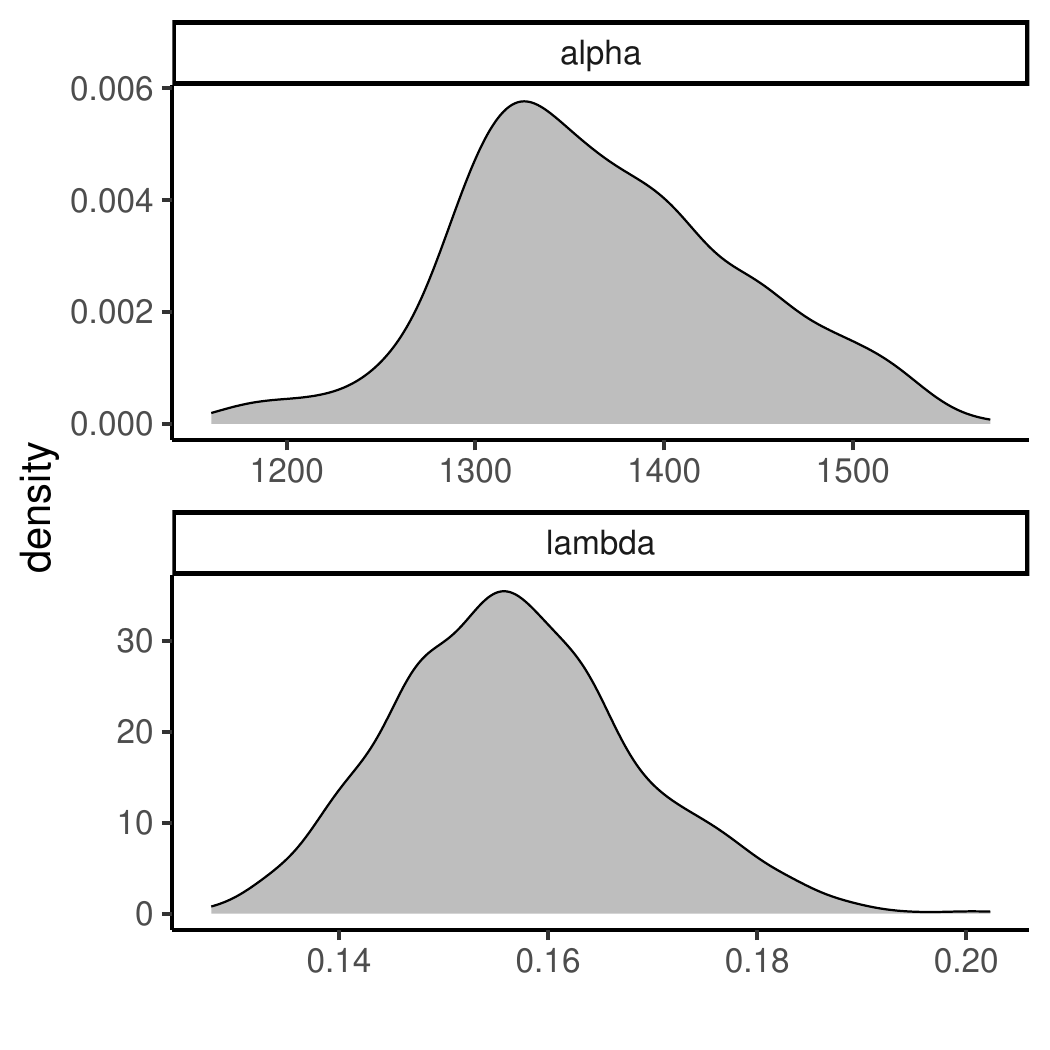}
      \label{sub:post_plot_Pois-Gamma}
                         }

    \caption{Posterior distribution of the parameters of the loss models when fitted to the pet insurance dataset.}
    \label{fig:post_plot_real}
  \end{center}
\end{figure}

We study $90$ quotes from $5$ insurers operating in the pet insurance market for a specific risk class associated to a 4-year-old female Australian Shepherd. We need to make some parametric assumptions to model claim frequency and severity. Classical claim frequency distributions include the Poisson, Binomial, and Negative Binomial distributions, which allow us to accommodate equidispersion, underdispersion, and overdispersion, respectively. For claim severity, we have chosen the Gamma and Lognormal distributions. The Gamma distribution is a common choice for modeling claim severity when using generalized linear models, but it is characterized by a light tail. The Lognormal distribution has a thicker tail, making larger claim sizes more likely to occur. We limit ourselves to two-parameter models: one parameter for claim frequency and another for claim severity. So we take the \(\PoissonDist(\lambda)\) to model the claim frequency\footnote{Note that the \texttt{IsoPriceR} package accomodate also the binomial and negative binomial distributions.} with the following prior setting:
\begin{equation}\label{eq:prior_settings_claim_frequency}
\lambda \sim\UnifDist([0,10]).
\end{equation}
We consider three claim severity distributions including \(\LognormalDist(\mu =0, \sigma)\), \(\LognormalDist(\mu, \sigma = 1)\), and \(\GammaDist(\alpha, \beta = 1)\). The prior settings over the parameters of the claim size distributions are as follows:
\begin{equation}\label{eq:prior_settings_claim_severity}
\mu \sim \UnifDist([-10,10]), \quad \sigma \sim \UnifDist([0,10]), \quad \text{and} \quad \alpha \sim \UnifDist([0, 10^5]).
\end{equation}
Combining the distributions for the claim frequency and severities results in a total of $3$ loss models. The population size in the \abc algorithm is set to \(J = 1{,}000\). The pure premiums are computed using \(R = 2{,}000\) Monte Carlo replications. The algorithm stops whenever the difference between two consecutive tolerance levels is lower than \(\Delta_\epsilon = 1\). The bounds for the loss ratio corridor are set to \(\text{LR}_\text{low} = 40\%\) and \(\text{LR}_\text{high} = 70\%\). The posterior distributions of the parameters for each model are provided in \cref{fig:post_plot_real}.
\newpage

For all the models, the algorithm updates the prior distribution in an informative way. \cref{tab:tolerance_level} provides the tolerance levels (ranked in increasing order) during the last iteration of the \abc algorithm for the loss models.  
\begin{table}[ht!]
\centering
\footnotesize
\begin{tabular}{lr}
  \toprule
Model & $\epsilon$ \\ 
  \midrule
  $\PoissonDist(\lambda)-\GammaDist(\alpha,\beta=1) $& 93.72 \\ 
  $\PoissonDist(\lambda)-\LognormalDist(\mu,\sigma=1) $& 96.20 \\ 
  $\PoissonDist(\lambda)-\LognormalDist(\mu=0,\sigma) $& 113.15 \\ 
   \bottomrule
\end{tabular}
\caption{Tolerance level during the last iteration of the \abc algorithm fo each loss model}
\label{tab:tolerance_level}
\end{table}

The final tolerance levels lies between $93.72$ and $113.15$ which is higher than the tolerance obtained in the simulation study which was around $33$ for $50$ data points and $50$ for $200$ data points. This discrepancy indicates misspecifications which stem from our assumptions about insurance companies adhering to the expectation principle for premium calculation and the models employed for claim frequency and claim amounts. \cref{tab:map_bp_real_data} reports the estimations of the parameters of all the model using the \map and the \mode.

 \begin{table}[ht!]
\centering
\footnotesize
\begin{tabular}{ll crr}
  \toprule
Model & &\phantom{abc} &\map & \mode \\ 
  \midrule
$\PoissonDist(\lambda)-\LognormalDist(\mu,\sigma=1)$ & $\lambda$ && 0.31 & 0.30 \\ 
   & $\mu$ && 6.14 & 6.19 \\ 
  $\PoissonDist(\lambda)-\GammaDist(\alpha,\beta=1)$ & $\lambda$ && 0.16 & 0.14 \\ 
  & $\alpha$ && 1365.53 & 1491.75 \\ 
  $\PoissonDist(\lambda)-\LognormalDist(\mu=0,\sigma)$ & $\lambda$ && 5.30 & 4.24 \\ 
  & $\sigma$ && 2.97 & 3.11 \\ 
   \bottomrule
\end{tabular}
\caption{\map and \mode estimator for the parameters of the loss models.}
\label{tab:map_bp_real_data}
\end{table}

\cref{tab:map_bp_Total_LR_severity_real_data} reports the estimations of the average total claim amounts and the average loss ratio for all the models for all models when fitted using the \map and the \mode.

\begin{table}[ht!]
\centering
\footnotesize
\begin{tabular}{lrrrrr}
  \toprule
&  \multicolumn{2}{c}{Loss ratio}  &\phantom{abc}& \multicolumn{2}{c}{$\mathbb{E}(X)$}\\
\cmidrule{2-3} \cmidrule{5-6}
Model & \map & \mode && \map & \mode \\ 
  \midrule
$\PoissonDist(\lambda)-\LognormalDist(\mu,\sigma=1)$ & 0.62 & 0.62 && 239.34 & 245.04 \\ 
  $\PoissonDist(\lambda)-\GammaDist(\alpha,\beta=1)$ & 0.66 & 0.60 && 225.36 & 208.23 \\ 
  $\PoissonDist(\lambda)-\LognormalDist(\mu=0,\sigma)$ & 0.62 & 0.61 && 422.78 & 730.47 \\ 
   \bottomrule
\end{tabular}
\caption{\map and \mode estimators of the average loss ratio and average total claim amounts.}
\label{tab:map_bp_Total_LR_severity_real_data}
\end{table}

\begin{figure}[ht!]
\centering
  \includegraphics[width=0.6\linewidth]{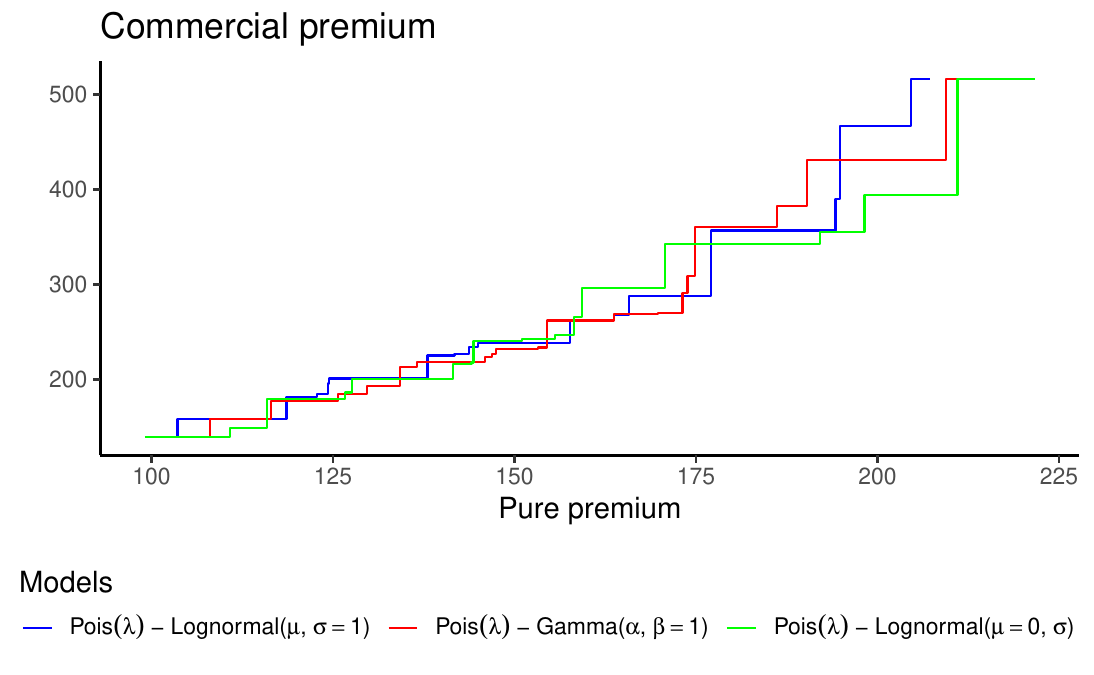}
  \caption{Isotonic link between pure and commercial premium for the different loss models.}
  \label{fig:iso_real_data_all_models}
\end{figure}

We note that the risk level, characterized here by the expected total claim amount, is similar for all the models, maybe a bit higher for the model having the $\LognormalDist(\mu = 0,\sigma)$ as claim sizes distribution. We further look into the loading function approximated via the isotonic regression. We estimate the pure premium for each model using the \map as an estimator of the model parameters and we plot the isotonic regression function to explain the commercial premium on \cref{fig:iso_real_data_all_models}.

\begin{figure}[ht!]
\centering
  \includegraphics[width=0.6\linewidth]{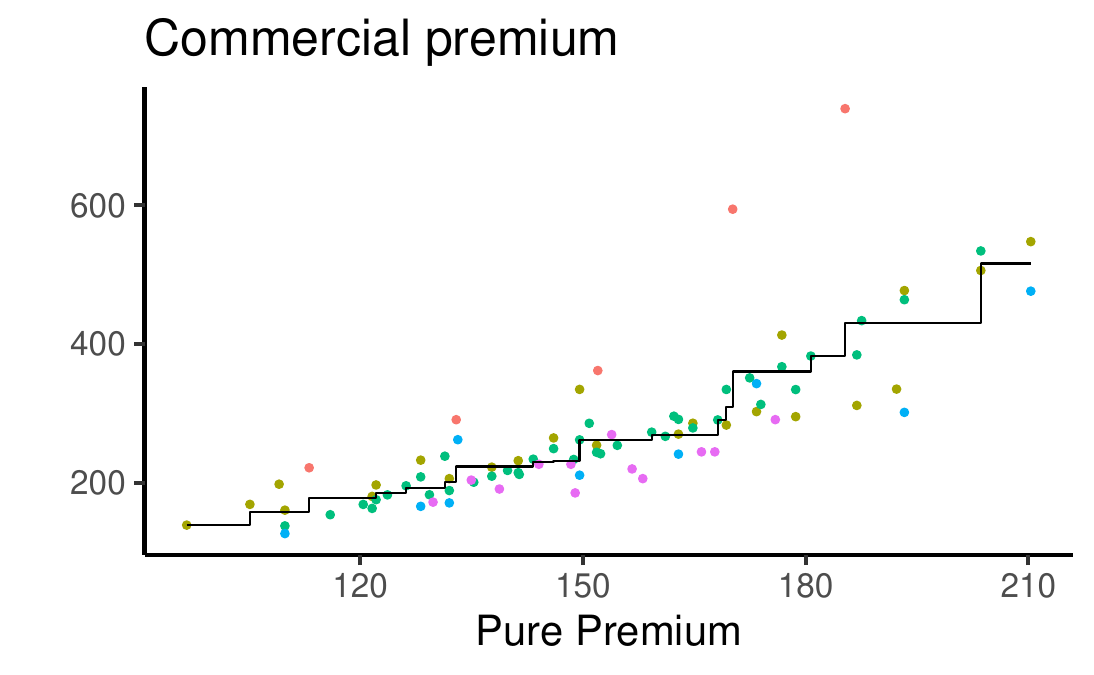}
  \caption{Commercial premium as a function of the pure premium for the $\PoissonDist(\lambda)-\LognormalDist(\mu,\sigma=1)$ depending on the insurance carrier.}
  \label{fig:iso_pois_lnorm_sigma1}
\end{figure}

The isotonic fits of the loading function accross all the models are similar which means that the models all agree on a common ordering of the pure premiums of the various insurance coverages. To highlight the explanatory power of our methodology, let's focus on the $\PoissonDist(\lambda)-\LognormalDist(\mu,\sigma=1)$ loss model. Note that the choice of the loss model is somewhat arbitrary because the information extracted from the data in \cref{fig:iso_real_data_all_models} is relatively consistent across the considered models. In \cref{fig:iso_pois_lnorm_sigma1}, we present a plot that illustrates the relationship between the commercial premium and the pure premium for the $\PoissonDist(\lambda)-\LognormalDist(\mu,\sigma=1)$ model. Different insurance companies are indicated by distinct colors, providing a visual representation of each company's respective rates. 

The accuracy of the loss model fitting enables us to condense the three-dimensional information of the rate of coverage, deductible, and limit into a single metric: the pure premium. Subsequently, isotonic regression unveils the relationship between commercial and pure premiums, providing a link between the two. The distinctions among various players in the pet insurance market come to light through the color-coded points, offering insights into the pricing strategies adopted by industry participants.

\subsection{Analysis of several risk classes with one model}\label{ssec:several_risk_classes_one_model}
The  $\PoissonDist(\lambda)-\LognormalDist(\mu,\sigma=1)$ model is fitted to the data within each risk classes (90 quotes) of \cref{tab:risk_classes}. The prior settings are given by 
$$
\lambda \sim\UnifDist([0,10])  \text{, and }\mu\sim\UnifDist([-10,10]).
$$
The algorithm's hyperparameters are similar to that of the previous subsection with 
\[
J = 1000, \text{ }R = 2 000, \text{ and }\Delta_\epsilon = 1.
\]

\begin{figure}[ht!]
  \begin{center}
    
    \subfloat[Posterior predictive distribution of $\mathbb{E}(X)$]{
      \includegraphics[width=0.4\textwidth]{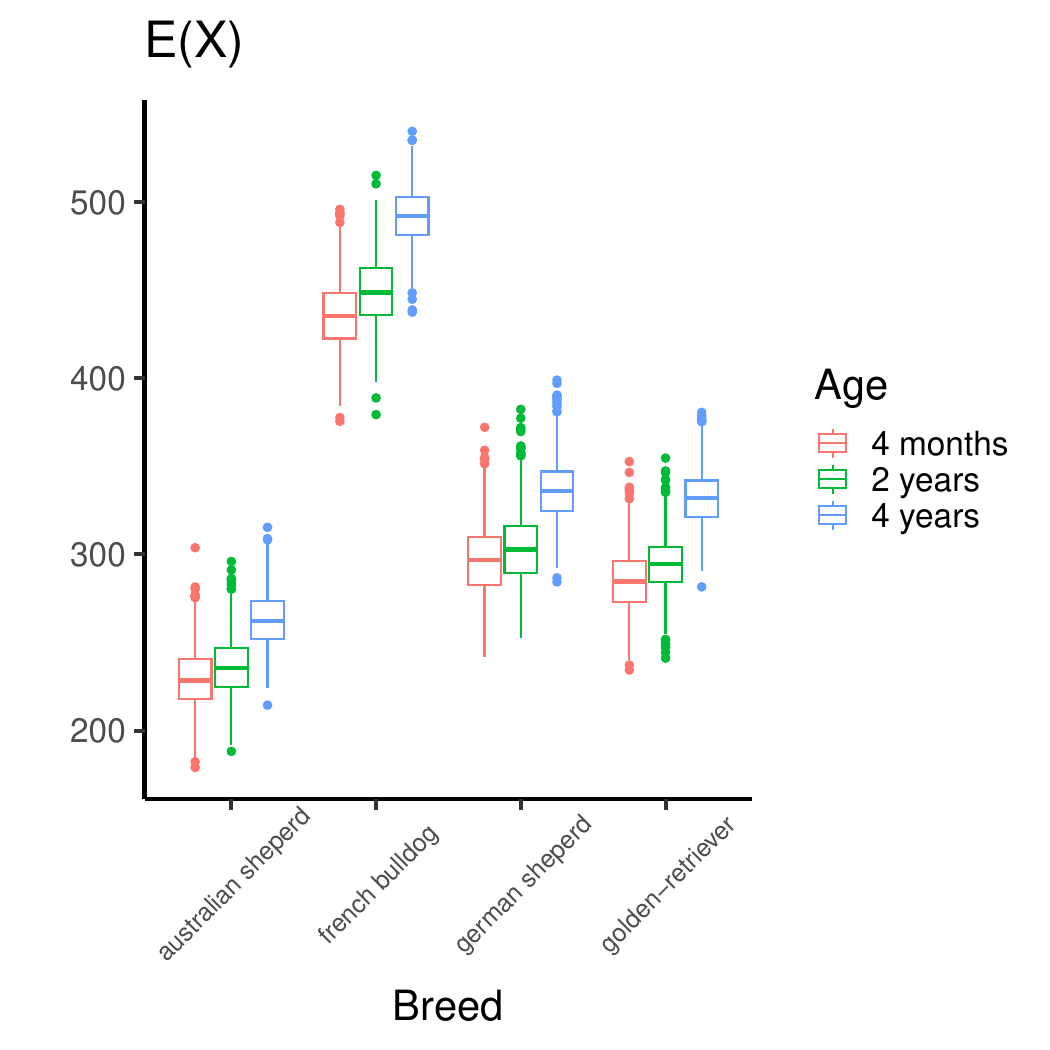}
      \label{sub:boxplot_EX_multiple_risk_classes}
                         }
    \subfloat[Posterior predictive distribution of the average loss ratio]{
      \includegraphics[width=0.4\textwidth]{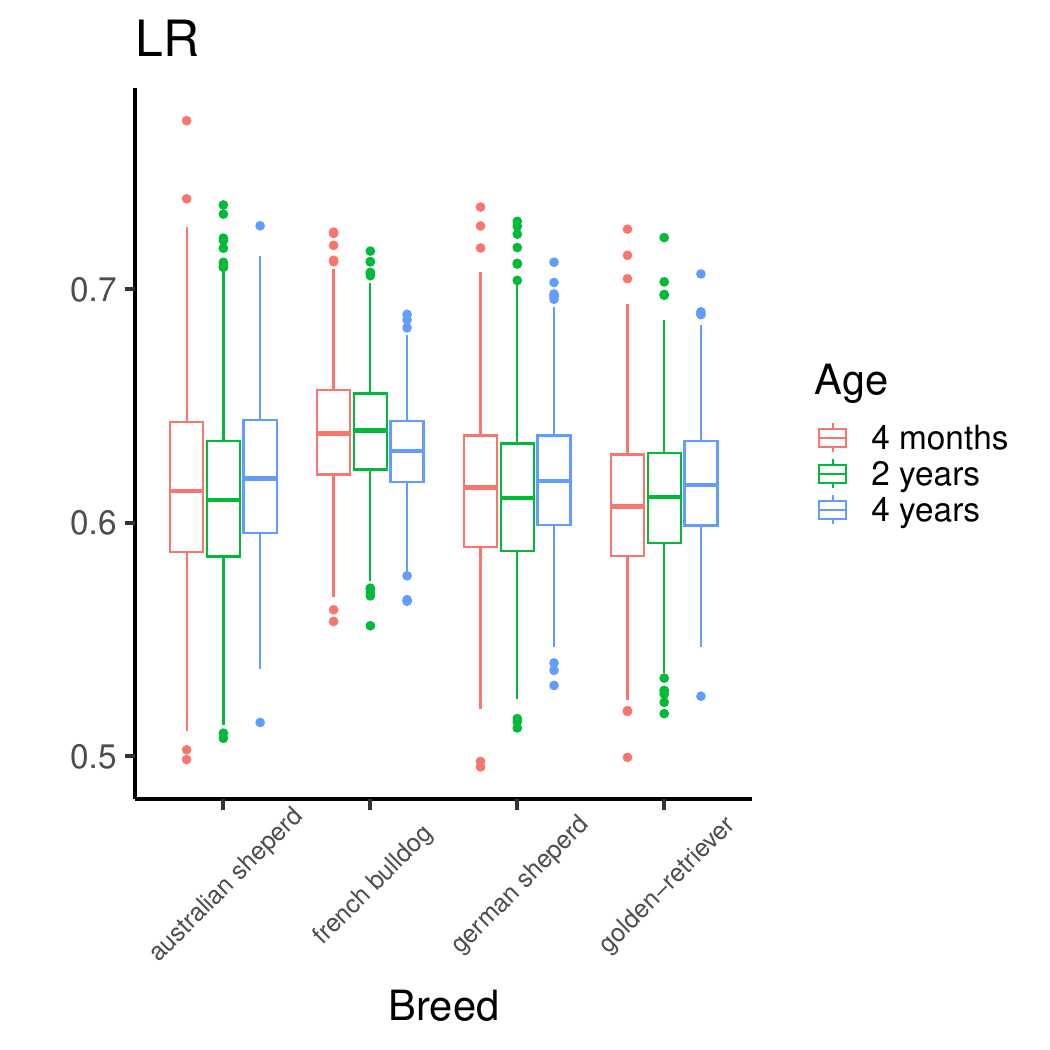}
      \label{sub:boxplot_LR_multiple_risk_classes}
                         }

    \caption{Posterior predictive distribution of $\mathbb{E}(X)$ and average loss ratio within each risk class.}
    \label{fig:post_boxplot_EX_LR_risk_class}
  \end{center}
\end{figure}

\cref{fig:post_boxplot_EX_LR_risk_class} shows the posterior predictive distribution of the expected total claim amounts and the averaged loss ratio within each risk class.

\cref{sub:boxplot_EX_multiple_risk_classes} allows us to compare the different risk classes. We note that an older dog is more expensive on average  and that the breeds may be ordered as Australian Sheperd, Golden-Retriever, German sheperd and french bulldog in terms of riskiness. \cref{sub:boxplot_LR_multiple_risk_classes} indicates that the loss ratios are arround $62-65\%$ for all the risk classes.

\section{Conclusion}%
\label{sec:conclusion}

We have developed a robust methodology for risk assessment based on market data  for pet insurance.  We employ a one-parameter model for the claim frequency and claim size distribution, connecting the pure premium to the commercial premium through an isotonic regression model. This approach optimizes the alignment between commercial and pure premiums while providing a framework for quantifying the associated parameter uncertainty through an Approximate Bayesian Computation algorithm.

The methodology's effectiveness and reliability have been validated within a simulation study and a practical application to an actual pet insurance dataset. This methodology is made accessible to the community through our R package, \texttt{IsoPriceR}\footnote{ see the \href{https://github.com/LaGauffre/market_based_insurance_ratemaking}{market\_based\_insurance\_ratemaking} \texttt{Github} repository}.



While this paper focuses on the specific context of pet insurance—a market that has experienced recent growth and significant capital investments—we believe that our methodology can be extended, with minor modifications, to other insurance products with straightforward compensation schemes. For example, this approach could be applied to unemployment benefits insurance, where the compensated risk would be redefined as follows: $x$ represents the number of days of actual unemployment, $r$ the daily compensation amount, while $d$ and $l$ continue to denote the deductible and guarantee limit, respectively.

However, our methodology may face limitations when applied to more complex insurance products and pricing structures, such as those commonly used in home or vehicle insurance. These products often rely on generalized linear models with high-dimensional covariates for tariff generation. In contrast, our research focuses on identifying only two parameters within a specific risk class. The latter may be based on several categorical variables. Commercial quotes will be needed within each risk classes making the data colllection impossible without a proper automation of the process. Extending the problem to a high-dimensional parameter space, such as 30 dimensions, would present significant challenges. Further research is needed to assess the feasibility of applying our core concept—leveraging commercial quotes to retro-engineer risk components—in such settings.

While we observe empirical convergence in our specific case, this work does not provide theoretical guarantees of convergence or an estimation of the data volume required to achieve it.

Finally, this paper primarily addresses risk assessment prior to the launch of a pet insurance product. Once launched, the company will begin collecting individual-level data. To improve pricing accuracy, the integration of this historical data should be considered as it becomes available. Consequently, a promising avenue for future research lies in developing a credibility framework that combines historical and market data, offering a comprehensive approach to risk assessment and pricing in emerging markets.


\section*{Acknowledgements}
Pierre-O's work is conducted within the Research Chair \href{https://chaire-dialog.fr/}{DIALOG} under the aegis of the Risk Foundation, an initiative by CNP Assurances. His research is also supported by the ANR project \href{https://anr.fr/en/funded-projects-and-impact/funded-projects/project/funded/project/b2d9d3668f92a3b9fbbf7866072501ef-7c3a1bbf87/?tx_anrprojects_funded%5Bcontroller%5D=Funded&cHash=e7c1441f5813b83b79e312bc2c3256a8}{DREAMES}.

\appendix
\section{Comparison of linear and isotonic regression to predict commercial premiums}\label{app:comparison_linear_isotonic}
\begin{figure}[!ht]
\centering
  \includegraphics[width=0.45\linewidth]{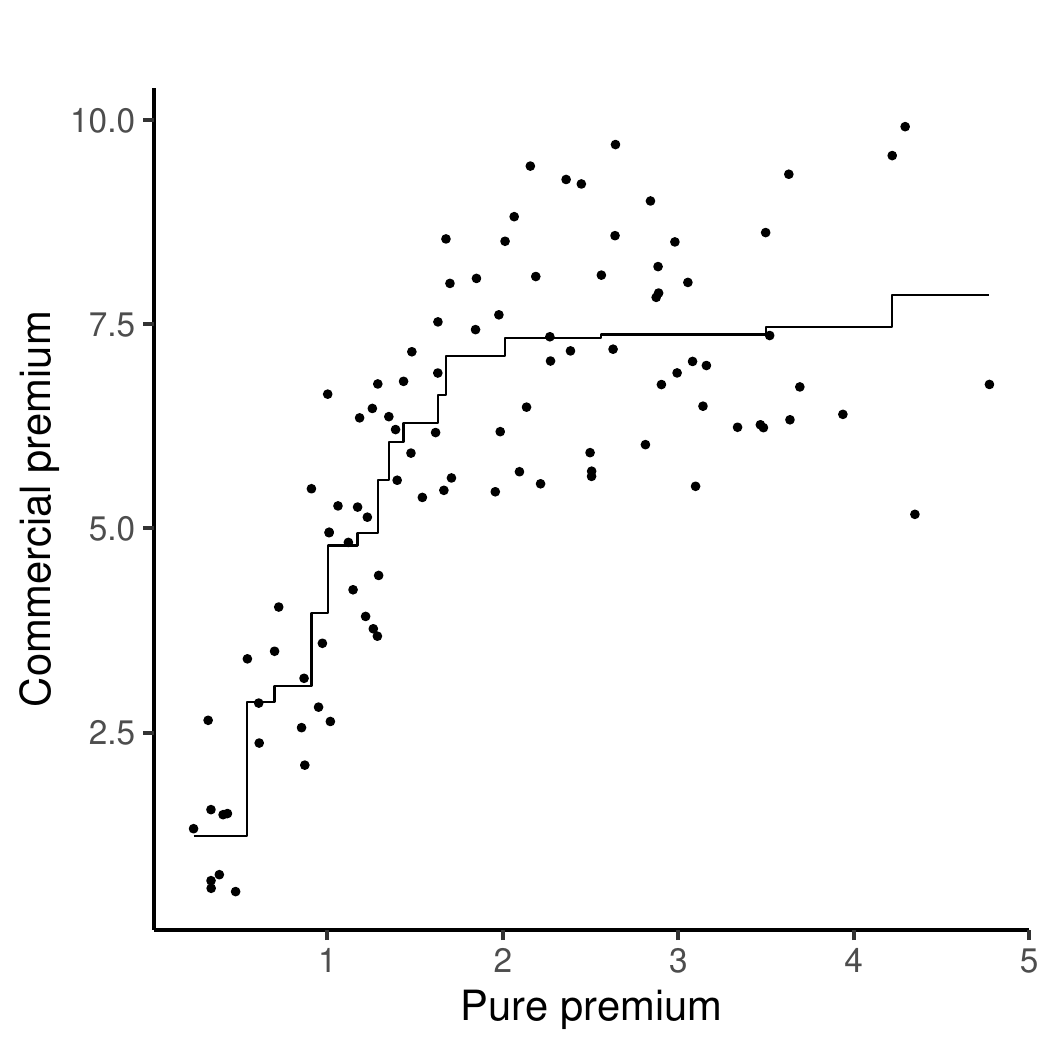}
  \caption{Isotonic link between the pure and commercial premiums.}
  \label{fig:gompertz_iso_plot}
\end{figure}

Instead of the linear link of \eqref{eq:cps} between pure and commercial premium, we consider 
\begin{equation*}
\tilde p_i = a_i\exp\left[-b_i\exp(c\cdot p_i)\right] ,\text{ for }i=1,\ldots, n. 
\end{equation*}
where 
$$
a_i \sim\UnifDist(5, 10]),\text{ }b_i \sim\UnifDist(2, 6]),\text{ and }c = 2\text{ for }i=1,\ldots, n. 
$$ 
This is a Gompertz growth curve type of link. The commercial premiums as a function of the pure premium is shown on \cref{fig:gompertz_iso_plot}.

\begin{figure}[!ht]
\centering
  \includegraphics[width=0.4\linewidth]{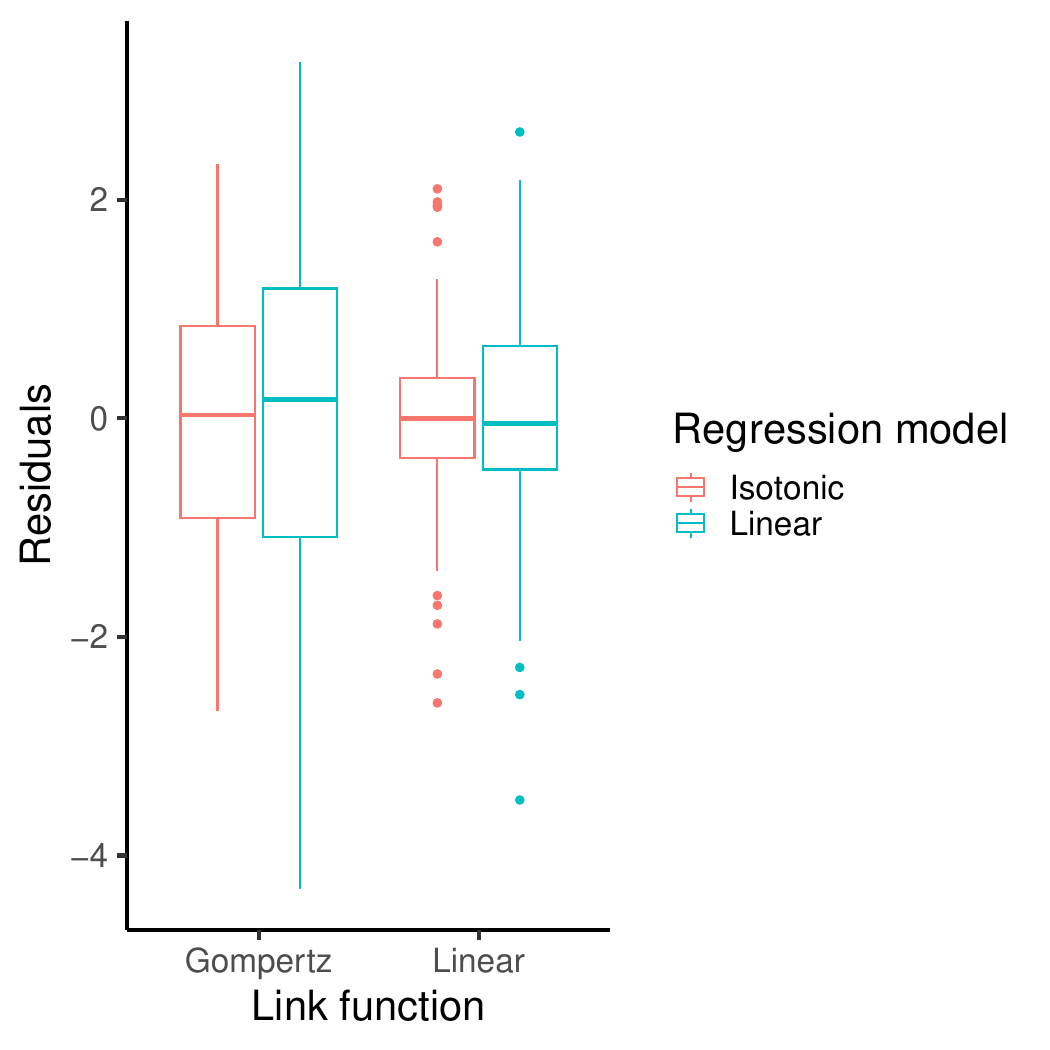}
  \caption{Boxplot of the residuals of the linear and isotonic regression models fo a linear and a Gompertz type link between pure and commercial premiums.}
  \label{fig:boxplot_residuals}
\end{figure}

We further compare the residuals of the isotonic regression model fitted to the data of \cref{fig:iso_plot,fig:gompertz_iso_plot} to that of a linear regression model fitted to the same data on \cref{fig:boxplot_residuals}
We note the proximity of the two models when the link between the pure and commercial premium is linear. When the link is not linear then isotonic regression model outperforms linear regression.

\section{Other premium principle}\label{app:other_premium_principle}
This paper focuses on the expectation premium principle as we try to inform the link $f$ between the commercial premium $\tilde p$ and the pure premium $p=\mathbb{E}[g(X)]$. Other premium principles such as the standard deviation principle can be considered by slightly adapting the method. Under such principle we have 
\begin{equation}\label{eq:std_principle}
\tilde p = f\left(\mathbb{E}[g(X)], \sqrt{\mathbb{V}[g(X)]}\right).
\end{equation}
where $f:\mathbb{R}_+\times\mathbb{R}_+\mapsto \mathbb{R}_+$. The same methodology applies, we simply need a bivariate model for $f$. The commercial premium should be increasing whenever the pure premium or the variance of the risk increases which leads to consider generalization of the univariate isotonic regression models which are readily available in the literature see the work of \citet{SASABUCHI1983}. More sophisticated premium principles such as the Escher principle or the utility indifference principle are also possible. Premium principles are described at length in actuarial science textbooks such as \citet{2005}. Considering a premium principle instead of another leads to model misspecification and will impact the final estimates of the underlying risk. 


\end{document}